\documentclass[aps,superscriptaddress,nofootinbib]{revtex4}
\usepackage{hyperref}
\usepackage{epsfig,rotating}
\usepackage{amsmath,amssymb}
\usepackage{dsfont}

\numberwithin{equation}{section}

\newcommand{\be}{\begin{equation}}
\newcommand{\ee}{\end{equation}}
\newcommand{\bea}{\begin{eqnarray}}
\newcommand{\eea}{\end{eqnarray}}

\newcommand{\tilqp}{\tilde{q}_+}
\newcommand{\tilqm}{\tilde{q}_-}
\newcommand{\tilsigpp}{\widetilde{\Sigma}_{++}}
\newcommand{\tilsigmm}{\widetilde{\Sigma}_{--}}
\newcommand{\tilsigpm}{\widetilde{\Sigma}_{+-}}
\newcommand{\tilsigmp}{\widetilde{\Sigma}_{-+}}

\begin{document}

\title{ Coherence and entanglement of mechanical oscillators mediated by \\ coupling   to different baths. }

\author{Daniel Boyanovsky}
\email{boyan@pitt.edu} \affiliation{Department of Physics and
Astronomy, University of Pittsburgh, Pittsburgh, PA 15260}
\author{David Jasnow}
\email{jasnow@pitt.edu}
\affiliation{Department of Physics and
Astronomy, University of Pittsburgh, Pittsburgh, PA 15260}
\date{\today}

\begin{abstract}
We study the non-equilibrium dynamics of two  mechanical oscillators with general linear couplings to two uncorrelated  thermal baths at   temperatures $T_1$ and $T_2$, respectively. We obtain the complete solution of the Heisenberg-Langevin equations, which reveal a coherent mixing among the normal modes of the oscillators as a consequence of their off-diagonal couplings to the baths. Unique  renormalization aspects resulting from  this mixing are discussed. Diagonal and off-diagonal (coherence) correlation functions are obtained analytically in the case of strictly Ohmic baths with different couplings in the strong and weak coupling regimes. An asymptotic non-equilibrium stationary state emerges for which we obtain the complete expressions for the correlations and coherence. Remarkably, the coherence survives in the high temperature, classical limit for $T_1 \neq T_2$.  This is a consequence of the coherence being determined by the \emph{difference} of the bath correlation functions.    In the case of vanishing detuning between the oscillator normal modes both coupling  to   one and the same bath, the  coherence   retains memory of the initial conditions at long time. A perturbative expansion of the early time evolution reveals that the emergence of  coherence is a consequence of the entanglement between the normal modes of the  oscillators \emph{mediated} by their couplings to the baths. This \emph{suggests} the survival of entanglement in the high temperature limit for different spectral densities and temperatures of the baths and is essentially  a consequence of the non-equilibrium nature of the asymptotic stationary state. An out of equilibrium setup with small detuning and    large $|T_1- T_2|$ produces    non-vanishing steady-state  coherence and entanglement in the high temperature limit of the baths.

\end{abstract}


\maketitle

\section{Introduction, motivation and goals}\label{sec:intro}
Progress in quantum information and quantum computing   with continuous   variables is receiving much attention with platforms that implement quantum optics, cavity quantum electrodynamics\cite{braun,girvin} and    optomechanics\cite{opto1,opto2,opto3,opto4,kipp}. The  quantum mechanical degrees of freedom  unavoidably couple to other environmental degrees of freedom, or bath, that induce dissipation and decoherence. Therefore, their dynamics    are treated as a  quantum open system\cite{breuer,weiss,gardiner,ines,alonso,marzo}. The motivation for a fundamental study of dissipation and decoherence by environmental degrees of freedom is bolstered by the theoretical\cite{paavo,brum,brunner,xu} and experimental\cite{bathengi,ver}   potential   to engineer the properties of the environmental bath. Quantum brownian motion\cite{feyver,leggett,ford1,hu1,hu2osc,flehu,breuer,weiss} provides a paradigmatic model to study the dynamics of open quantum systems.   In this model a quantum mechanical oscillator is linearly coupled to a bath described also by a (large) number of quantum mechanical oscillators; the properties of this bath are determined by its spectral density. This simple yet illuminating model has yielded a deep understanding of the role of an  environmental bath on decoherence and dissipation of quantum mechanical degrees of freedom.

This model also provides a theoretical foundation  for the emerging field of quantum thermodynamics\cite{quthermo1,quthermo2,quthermo3,paz,buttner1,kosloff} and quantum entanglement induced by coupling to environmental degrees of freedom\cite{buttner,marq,ronca,estrada}.

A remarkable experiment\cite{groex} that uses an opto-mechanical resonator  probed the spectral properties of an environmental bath coupled to the micro-mechanical oscillators, thus paving the way towards a deeper understanding of the effects of the coupling between quantum mechanical and  environmental degrees of freedom in experimentally controlled platforms.

The possibility of engineering the properties of the environmental degrees of freedom including coupling to several different  baths,   tailoring   decoherence and dissipative properties could allow novel cooling techniques of optomechanical systems\cite{xu}. Recent experimental studies\cite{exp1,exp2,exp3} have demonstrated the feasibility of coupling various quantum systems to   non-equilibrium baths, and of tailoring the environmental degrees of freedom. Recent theoretical analysis\cite{brum}    of a V-type   molecular system showed the possibility of long-lived coherences despite     noise-induced decoherence, and ref.\cite{brunner} reports the emergence of  steady-state entanglement in the case of finite dimensional coupled systems such as superconducting qubits or quantum dots. Furthermore, coupling to different non-equilibrium  environments has been argued to lead to the persistence of entanglement in the high temperature limit\cite{hiTent,estrada,hiTvedral} as well as to the coherent mixing of mechanical excitations in nano-optomechanical structures\cite{painter}. Ref.\cite{marq} establishes bounds on the couplings to the baths to maximize entanglement of the quantum degrees of freedom induced by these couplings. A  widely used approach to studying decoherence and dissipation relies on the quantum master equation for the reduced density matrix of the system, either in the form of influence functionals obtained in ref.\cite{hu1} or of the Lindblad form\cite{breuer,weiss,gardiner}   after tracing over the bath degrees of freedom. Recent articles\cite{joshi,keeling}  study the dynamics of quantum coherence for a   system of two oscillators coupled  with different couplings  to a thermal bath  within the framework of the quantum master equation. These studies question the applicability  of the Lindblad quantum master equation approach to the non-equilibrium dynamics and suggest     important modifications  in the case of a system of several oscillators coupled to a bath.

\vspace{2mm}

\textbf{Motivation and goals:}

Experimental advances in cavity electrodynamics and optomechanics along with the possibility of engineering the spectral properties of environmental baths   open  new avenues for quantum information and quantum computing and provide new platforms for  fundamental studies of coherence,  dissipation and quantum thermodynamics.  Motivated by these developments  we   study the non-equilibrium dynamics of two coupled ``mechanical'' oscillators, each one in turn coupled   to two different and uncorrelated baths at different temperatures. This model is a generalization of those considered  in refs.\cite{buttner,marq,xu,estrada,keeling,painter} which mostly focused on the influence functional\cite{hu1,paz,estrada} or Lindblad approach to the quantum master equation\cite{breuer,gardiner}. Instead,  we solve exactly the Heisenberg-Langevin equations for general couplings and spectral densities of the baths. Additionally, we focus   specifically on the Ohmic case, which allows an analytic treatment,  addressing in detail correlations and coherences in the asymptotic stationary regime.

Our main \emph{goals} are:

\begin{itemize}
\item{ To provide a general solution of the Heisenberg-Langevin equations valid for two mechanical oscillators with   linear couplings to different environments with general spectral densities. We study the non-equilibrium dynamics as an initial value problem to reveal the approach to a stationary regime in the asymptotic long time limit, allowing one to deal with cases in which correlation functions  retain  memory of the initial conditions in this limit. }

\item{ The  provide a complete analytic treatment in the case of Ohmic baths. In this exactly solvable case we study the emergence of an asymptotic stationary but non-equilibrium state. We focus on two relevant cases: a strong coupling case that includes   vanishing detuning between the normal modes of the mechanical oscillators, and a weak coupling case corresponding to weak damping on the scale of the normal mode frequencies.}

\item{ To obtain the  correlation functions and coherence (off diagonal correlations between the normal modes) at different times. Specifically we focus on the interesting and potentially observable survival of coherence in the classical high temperature limit.  }

\item{ To discuss renormalization aspects that arise from the coupling of  individual oscillators to two different baths, which results in ``mixing'' of the normal modes of the mechanical oscillators, requiring new renormalization counterterms. }

  \item{ To explore the temporal emergence of entanglement and coherence between the normal modes   \emph{mediated} by their couplings to and entanglement with the bath degrees of freedom.}
\end{itemize}

\vspace{2mm}

\textbf{Brief summary of results:}

\begin{itemize}
\item{We obtain the general solution of the Heisenberg-Langevin equation for the case of two mechanical oscillators with   linear couplings to two different and uncorrelated baths. The different couplings to the baths lead to a ``mixing'' of the normal modes of the mechanical oscillators. This mixing implies novel renormalization aspects and the necessity for off-diagonal counterterms in the effective Hamiltonian of the mechanical oscillators.  }

    \item{The case of strictly Ohmic baths\cite{breuer,weiss} yields an analytic solution for the time evolution of the correlation functions. This solution allows one to obtain correlation functions and coherence (off diagonal correlations of the mechanical normal modes)  at different times without the need to invoke the quantum regression theorem\cite{breuer,gardiner}.  A stationary non-equilibrium state emerges in the asymptotic long time-limit. We find both in the strong and weak coupling regimes that coherence survives in the high temperature limit if the baths feature \emph{different spectral densities and/or temperatures}.


        We show that this   surprising and counterintuitive result is a consequence of  the coherence being  proportional to the \emph{difference} of the correlation functions of the baths degrees of freedom.

         In the strong coupling case coherence survives in the high temperature limit if the baths feature different temperature, whereas in the weak coupling case coherence survives in this limit if   the temperature and/or the couplings are different.   Classical equilibrium equipartition follows in both cases   when the baths are at the same temperature; in this equilibrium case  the coherence is suppressed in the weak coupling case. }

        \item{When the renormalized normal modes of the mechanical oscillators are degenerate and both couple only to one bath, we find that the asymptotic long time limit retains memory of the initial conditions, possibly suggesting a breakdown of a Markovian approximation.}

\item{ A perturbative expansion is implemented to learn about the early time evolution of the coherence in the case when both baths are at zero temperature. We find that the emergence of coherence is a manifestation of the entanglement between the normal modes of the mechanical oscillators \emph{mediated} by their entanglement with the environmental degrees of freedom. This study leads us to \emph{conjecture} that  entanglement  may survive in the high temperature limit (of each bath) when the baths are at different temperatures with large $|T_1-T_2|$  relative to the frequencies of the normal modes.  We conclude that this result opens the possibility of designing a setup  of mechanical oscillators coupled to two baths at different temperatures as  a platform that maintains coherence and entanglement in the high temperature limit of both baths.  }

\end{itemize}

\section{The model and the general solution of the equations of motion. }\label{sec:model}
We consider two interacting oscillators of equal unit mass coupled to two different baths in equilibrium at different temperatures, specifically
\be H= H_S+ H_B+ H_{SB} \label{totalH}\ee
with
\be H_S = \frac{p^2_a}{2}+\frac{p^2_b}{2}+\frac{\Omega^2_a}{2}~q^2_a + \frac{\Omega^2_b}{2}~q^2_b+\frac{\Omega^2}{2}~(q_a-q_b)^2\,, \label{Hsis}\ee and

\be H_B = \sum_{p} \Big[\frac{P^2_{p,1}}{2}+\frac{W^2_{p,1}}{2}\, Q^2_{p,1}\Big] + \sum_{k} \Big[\frac{P^2_{k,2}}{2}+\frac{W^2_{k,2}}{2}\, Q^2_{k,2}   \Big]\,. \label{Hbath}\ee The  Hamiltonian $H_B$ describes two independent baths of harmonic oscillators to be taken in thermal equilibrium at different temperatures $T_1,T_2$ respectively. The system-bath coupling is taken to be

\be  H_{SB}   =   -(q_a \cos(\theta) + q_b \sin(\theta))\,B_1[\{Q_1\}] -(q_b \cos(\theta) - q_a \sin(\theta))\,B_2[\{Q_2\}] \label{SBcoup} \ee where $\theta$ is an arbitrary mixing angle and
\be B_1[\{Q_1\}]   =   \sum_{p} C_p \,Q_{p,1}  ~~;~~
B_2[\{Q_2\}]   =   \sum_{k} D_k \,Q_{k,2}\,. \label{baths} \ee

This system-bath Hamiltonian generalizes the cases studied in refs.\cite{keeling,marq}, and as   will be seen below, it describes mixing and coherence among the mechanical oscillators similar to  the coherent mixing of excitations in nano-optomechanical structures described  in ref.\cite{painter} and the case of two-mode coupling in multimode cavity quantum optomechanics\cite{meystre}. In ref.\cite{keeling} the case $\Omega=0$ was considered with only one bath; after phase redefinitions  the system bath coupling was taken as $(q_a \, \phi_a + q_b\,\phi_b) B[\{Q_1\}]$. Writing $\phi_a = \sqrt{\phi^2_a+\phi^2_b}\,\cos(\theta)\,,\, \phi_b = \sqrt{\phi^2_a+\phi^2_b}\,\sin(\theta)$ and absorbing $\sqrt{\phi^2_a+\phi^2_b}$ into a redefinition of the couplings $C_p$ in the system-bath Hamiltonian (\ref{SBcoup},\ref{baths}), the model studied in ref.\cite{keeling} is equivalent to the generalized model described above with $\Omega=0$ and $D_k =0 \,\forall \, k$.

It is convenient to introduce the vector
\be \vec{q}=  \Big( \begin{array}{c}
                 q_a \\
                 q_b
               \end{array}\Big) \label{vecqs}
\ee and the frequency matrix
\be \mathbf{\Omega}^2 = \left(
                        \begin{array}{cc}
                          \Omega^2_a+\Omega^2 & -\Omega^2 \\
                          -\Omega^2 & \Omega^2_b+ \Omega^2 \\
                        \end{array}
                      \right) = \frac{1}{2}\Big(\Omega^2_a + \Omega^2_b + 2\Omega^2 \Big) \mathds{1} + \frac{1}{2} \,\left(
                                      \begin{array}{cc}
                                        \Omega^2_a-\Omega^2_b & -2\Omega^2 \\
                                        -2\Omega^2 & \Omega^2_b-\Omega^2_a \\
                                      \end{array}
                                    \right)
\label{freqmtx} \ee

\noindent to write the potential term in $H_S$ as $ \vec{q}^{\,T}\,\mathbf{\Omega}^2\, \vec{q} $; diagonalizing this quadratic form we obtain the normal modes of the coupled oscillators. Introducing
\be   \cos(2\lambda) = \frac{(\Omega^2_a-\Omega^2_b)}{\sqrt{(\Omega^2_a-\Omega^2_b)^2+4\Omega^4}}~~;~~ \sin(2\lambda) = \frac{2\Omega^2}{\sqrt{(\Omega^2_a-\Omega^2_b)^2+4\Omega^4}} \label{defs} \ee the matrix $\mathbf{\Omega}^2 $ is written as
\be \mathbf{\Omega}^2 = \left(
                        \begin{array}{cc}
                          \Omega^2_a + \Omega^2& -\Omega^2 \\
                          -\Omega^2 & \Omega^2_b + \Omega^2 \\
                        \end{array}
                      \right) = \frac{1}{2}\Big(\Omega^2_a + \Omega^2_b + 2\Omega^2 \Big) \mathds{1} + \frac{1}{2}\,\sqrt{(\Omega^2_a-\Omega^2_b)^2+4\Omega^4} \,\,\left(
                                      \begin{array}{cc}
                                        \cos(2\lambda) & -\sin(2\lambda) \\
                                        -\sin(2\lambda) & -\cos(2\lambda) \\
                                      \end{array}
                                    \right)\,,
\label{mixangs} \ee which yields a straightforward diagonalization by a unitary transformation, namely
\be \left(
                                      \begin{array}{cc}
                                        \cos(2\lambda) & -\sin(2\lambda) \\
                                        -\sin(2\lambda) & -\cos(2\lambda) \\
                                      \end{array}
                                    \right) = \mathbf{V}^{-1}(\lambda) \, \left(
                                      \begin{array}{cc}
                                        1 & 0 \\
                                        0 & -1 \\
                                      \end{array}
                                    \right) \,   \mathbf{V}(\lambda) \label{diag} \ee with

\be \mathbf{V}(\lambda) =  \left(
                                      \begin{array}{cc}
                                        \cos(\lambda) & -\sin(\lambda) \\
                                         \sin(\lambda) & \cos(\lambda) \\
                                      \end{array}
                                    \right)\,. \label{Voflam}\ee Therefore
 \be \mathbf{\Omega}^2 = \mathbf{V}^{-1}(\lambda) \, \left(
                                                       \begin{array}{cc}
                                                         \Omega^2_+ & 0 \\
                                                         0 & \Omega^2_- \\
                                                       \end{array}
                                                     \right)
  \, \mathbf{V}(\lambda) \,, \label{Omemtxdiag}\ee
  where
  \be \Omega^2_\pm = \frac{1}{2}\,\Big(\Omega^2_a + \Omega^2_b + 2\Omega^2 \Big) \pm \frac{1}{2} \,\sqrt{(\Omega^2_a-\Omega^2_b)^2+4\Omega^4}  \label{NMfreqs} \ee are the normal mode frequencies.

                                    The coordinates and momenta of the normal modes ($q_\pm;p_\pm$), are related to the original ones  ($q_{a,b};p_{a,b}$) by
\be
\left(
             \begin{array}{c}
               q_+ \\
               q_- \\
             \end{array}
           \right)  = \mathbf{V}(\lambda) \, \left( \begin{array}{c}
               q_a \\
               q_b \\
             \end{array}
           \right) ~~;~~  \left(
             \begin{array}{c}
               p_+ \\
               p_- \\
             \end{array}
           \right)  = \mathbf{V}(\lambda) \, \left( \begin{array}{c}
               p_a \\
               p_b \\
             \end{array}
           \right) \,.\label{NMpq} \ee  The system Hamiltonian $H_S$ (\ref{Hsis}) becomes diagonal in the normal mode basis, the coordinates and momenta $q_\pm, p_\pm$ describing independent, uncoupled harmonic oscillators of frequencies $\Omega_\pm$ respectively.

           The system-bath coupling Hamiltonian (\ref{SBcoup},\ref{baths})) is written in the original basis as
 \be H_{SB} = -(q_a \, q_b) \,\left(
                               \begin{array}{cc}
                                 \cos(\theta) & -\sin(\theta) \\
                                 \sin(\theta) & \cos(\theta) \\
                               \end{array}
                             \right)\,\left(
                                        \begin{array}{c}
                                          B_1[\{Q_1\}] \\
                                          B_2[\{Q_2\}] \\
                                        \end{array}
                                      \right)\,,
  \label{SBmtx}\ee while in the normal mode basis ($(q_+\,q_-) = (q_a \, q_b)\,\mathbf{V}^{-1}(\lambda)$)   it becomes
   \be H_{SB} = -(q_+ \, q_-) \,\left(
                               \begin{array}{cc}
                                 \cos(\psi) & -\sin(\psi) \\
                                 \sin(\psi) & \cos(\psi) \\
                               \end{array}
                             \right)\,\left(
                                        \begin{array}{c}
                                          B_1[\{Q_1\}] \\
                                          B_2[\{Q_2\}] \\
                                        \end{array}
                                      \right)~~;~~ \psi = \lambda + \theta \,.
  \label{SBNM}\ee


  In the normal mode basis the total Hamiltonian is
  \be H_{tot}= \frac{p^2_+}{2} + \frac{\Omega^2_+}{2}\,q^2_+ + \frac{p^2_-}{2} + \frac{\Omega^2_-}{2}\,q^2_- + H_B + H_{SB} \,.\label{NMbasis} \ee where $H_{SB}$ is given by (\ref{SBNM}) in the normal mode basis.

  The main conclusion from this analysis is that the form of the total  Hamiltonian (\ref{NMbasis}) is \emph{general} even when the mechanical oscillators are coupled.


  The equations of motion in the normal mode basis are
  \bea
  \ddot{q}_+ + \Omega^2_+ q_+  & = &  \cos(\psi) \,\sum_{p} C_p \,Q_{p,1} - \sin(\psi) \,\sum_{k} D_k \,Q_{k,2}  \label{EOMplu} \\
   \ddot{q}_- + \Omega^2_- q_+  & = &  \sin(\psi)\, \sum_{p} C_p \,Q_{p,1} + \cos(\psi)\, \sum_{k} D_k \,Q_{k,2}  \label{EOMmin} \\
\ddot{Q}_{p,1} + W^2_{p,1} Q_{p,1} & = &  C_p\,\big[\cos(\psi)\,  q_+(t) + \sin(\psi)\, q_-(t)\Big] \label{EQMB1}\\
\ddot{Q}_{k,2} + W^2_{k,2} Q_{k,2} & = &  D_k\,\big[\cos(\psi)\,  q_-(t) - \sin(\psi)\, q_+(t)\Big] \,. \label{EQMB2}  \eea Treating the dynamics as an initial condition problem will allow us to examine the approach to a stationary  state  and correlations in this state in what follows.
We proceed to solve the equations of motion for the bath variables and insert the solution into the equations of motion for the system coordinates, namely
\bea Q_{p,1}(t) & = &  Q^{(0)}_{p,1}(t) + C_p \int_0^t \frac{\sin\big[W_{p,1}(t-t') \big]}{W_{p,1}} \Big(\cos(\psi)\,  q_+(t') + \sin(\psi)\, q_-(t') \Big)\,dt' \label{solB1}\\
 Q_{k,2}(t) & = &  Q^{(0)}_{k,2}(t) + D_k \int_0^t \frac{\sin\big[W_{k,2}(t-t') \big]}{W_{k,2}} \Big(\cos(\psi)\,  q_-(t') - \sin(\psi)\, q_+(t') \Big)\,dt' \,,\label{solB2} \eea where
 \bea Q^{(0)}_{p,1}(t) & = & \frac{1}{\sqrt{2\,W_{p,1}}}\,\Big[\alpha_{p}\,e^{-iW_{p,1}t}+\alpha^\dagger_p\,e^{ iW_{p,1}t}\Big] \label{zeroQ1}\\
 Q^{(0)}_{k,2} (t) & = & \frac{1}{\sqrt{2\,W_{k,2}}}\,\Big[\beta_{k}\,e^{-iW_{k,2}t}+\beta^\dagger_k\,e^{ iW_{k,2}t}\Big] \label{zeroQ2} \eea are the ``free-field'' operator solutions of the homogeneous equation in terms of independent  annihilation ($\alpha_p,\beta_k$) and creation ($\alpha^\dagger_p,\beta^\dagger_k$) bath operators. The independent baths are assumed in equilibrium at temperatures $T_1,T_2$ respectively, with statistical averages
 \be \langle \alpha^\dagger_p \alpha_p \rangle_1 = \frac{1}{e^{ W_{p,1}/T_1}-1}~~;~~
\langle \beta^\dagger_k \beta_k \rangle_2 = \frac{1}{e^{ W_{k,2}/T_2}-1} \,.\label{thermal}\ee Inserting the solutions (\ref{solB1},\ref{solB2}) into (\ref{EOMplu},\ref{EOMmin}) we find a system of \emph{coupled} Heisenberg-Langevin equations, namely
\bea
\ddot{q}_+(t) + \Omega^2_+ q_+(t) \, +\int_0^t \Big[\Sigma_{++}(t-t')q_+(t') + \Sigma_{+-}(t-t')q_-(t')\Big]\,dt'   & = & \xi_+(t) \label{HLplu} \\
\ddot{q}_-(t) + \Omega^2_- q_-(t) \, +\int_0^t  \Big[\Sigma_{--}(t-t')q_-(t') + \Sigma_{-+}(t-t')q_+(t')\Big]\,dt' & = & \xi_-(t)\,. \label{HLmin} \eea The self-energy kernels are given by
\bea \Sigma_{++}(t-t') & = &  \cos^2(\psi)\, \Sigma_1(t-t')+ \sin^2(\psi)\,\Sigma_2(t-t') \label{pluplu}\\
 \Sigma_{--}(t-t') & = &  \sin^2(\psi)\, \Sigma_1(t-t')+ \cos^2(\psi)\,\Sigma_2(t-t') \label{minmin} \\
  \Sigma_{+-}(t-t') & = &  \Sigma_{-+}(t-t') = \cos(\psi)\, \sin(\psi) \Big[\Sigma_1(t-t')-\Sigma_2(t-t')\Big] \,. \label{plumin} \eea Here the self-energies $\Sigma_{1,2}$ for each bath are given by

  \be  \Sigma_{1,2}(t-t')  =    \frac{i}{\pi}\,\int_{-\infty}^{\infty}  \sigma_{1,2}(\omega') \,e^{i\omega'(t-t')}\,d\omega' \, \label{sigmas}\ee  in terms of  the spectral densities of the baths   given by
  \bea \sigma_1(\omega') & = \sum_p \frac{\pi\,C^2_p}{2\,W_{p,1}}\,\big[\delta(\omega'-W_{p,1})-\delta(\omega'+W_{p,1}) \big] \label{spec1} \\
  \sigma_2(\omega') & = \sum_k \frac{\pi\,D^2_k}{2\,W_{k,2}}\,\big[\delta(\omega'-W_{k,2})-\delta(\omega'+W_{k,2}) \big] \label{spec2} \,, \eea   and the noise terms
  \be \left(
        \begin{array}{c}
          \xi_{+}(t) \\
          \xi_{-}(t) \\
        \end{array}
      \right)  =  \left(
                               \begin{array}{cc}
                                 \cos(\psi) & -\sin(\psi) \\
                                 \sin(\psi) & \cos(\psi) \\
                               \end{array}
                             \right) \, \left(
        \begin{array}{c}
          \xi_{1}(t) \\
          \xi_{2}(t) \\
        \end{array}
      \right)  = \mathbf{V}(\psi) \,\left(
        \begin{array}{c}
          \xi_{1}(t) \\
          \xi_{2}(t) \\
        \end{array}
      \right) \label{xis} \ee

 \noindent  where
  \be \xi_1(t)= \sum_p C_p Q^{(0)}_{p,1}(t)~~;~~ \xi_2(t)=\sum_k D_k Q^{(0)}_{k,2}(t)\,. \label{noises}\ee  We note that the spectral densities $\sigma_{1,2}(\omega')$ are \emph{odd} functions of $\omega'$.

  The thermal baths at temperatures $T_1,T_2$, respectively, are independent and uncorrelated, and  the noise correlation functions for each bath are obtained simply from (\ref{zeroQ1},\ref{zeroQ2},\ref{thermal}). We find
  \be  \langle\langle \xi_i(t)  \rangle \rangle =0 ~~;~~ \langle\langle \xi_i(t)\xi_j(t') \rangle \rangle   = \delta_{ij} ~ \frac{1}{\pi}\,\int_{-\infty}^{\infty} \sigma_{i}(\omega') \, n_i(\omega')  \, e^{i\omega'(t-t')}\, d\omega' ~~;~~ n_i(\omega)= \frac{1}{e^{\omega/T_i}-1} ~~;~~ i,j = 1,2 \label{correnoise} \ee
    where $\langle \langle (\cdots) \rangle \rangle$ correspond to statistical averages over the bath variables and $\sigma_{1,2}(\omega)$ are given by (\ref{spec1},\ref{spec2}), respectively. This relationship between the noise correlation functions and the self-energies   in  (\ref{sigmas}) is a manifestation of the fluctuation-dissipation relation independently for each bath.

    We note that as a consequence of the couplings to the baths, the equations of motion (\ref{HLplu},\ref{HLmin}) are \emph{off diagonal} in the normal mode basis; in other words the coupling to the baths induces a ``mixing'' between the normal modes. This mixing is at the heart of the bath-induced coherence\cite{keeling}, namely off diagonal correlations  between the normal modes  and, as   will be shown in a later section, of the entanglement between the normal modes.

  The Heisenberg-Langevin equations of motion (\ref{HLplu},\ref{HLmin}) are solved via Laplace transform, with
  \be \tilde{q}_{\pm}(s) = \int_0^\infty e^{-st}\,q_{\pm}(t)\,dt~~;~~ \widetilde{\Sigma}_{\alpha \beta }(s) =\int_0^\infty e^{-st}\,{\Sigma}_{\alpha \beta}(t)\,dt ~~;~~\tilde{\xi}_{\pm}(s) = \int_0^\infty e^{-st}\,{\xi}_{\pm}(t)\,dt \,, \label{laplavars}
  \ee with $\alpha,\beta = \pm $. From equation (\ref{sigmas}) it follows that
  \be \widetilde{\Sigma}_{1,2}(s) = - \int_{-\infty}^{\infty} \, \frac{\sigma_{1,2}(\omega')}{\omega'+ i\,s}\,\frac{d\omega'}{\pi} \,. \label{sigmaofs} \ee

 \noindent  In terms of the Laplace transforms  (\ref{HLplu},\ref{HLmin}) become
  \be \mathbf{G}^{-1}(s) ~ \left(
                             \begin{array}{c}
                               \tilqp(s) \\
                               \tilqm(s) \\
                             \end{array}
                           \right) = \left(
                                       \begin{array}{c}
                                         \tilde{J}_+(s) \\
                                         \tilde{J}_-(s) \\
                                       \end{array}
                                     \right) \ee  where

 \be   \mathbf{G}^{-1}(s) = \left(
                              \begin{array}{cc}
                                s^2+\Omega^2_+ + \tilsigpp(s) & \tilsigpm(s) \\
                                \tilsigmp(s) & s^2+\Omega^2_- + \tilsigmm(s) \\
                              \end{array}
                            \right) \label{invG}\ee and

 \be \tilde{J}_\pm(s) =   \dot{q}_\pm(0)+ s q_\pm(0) +\tilde{\xi}_\pm(s) \,.\label{Jpm} \ee

\noindent We note that the couplings to the bath induce \emph{off diagonal} entries in the matrix equations of motion; as discussed below these off-diagonal, bath-induced terms will lead to coherence between the different normal modes.

Furthermore, the analysis presented above shows that the coupling to the baths induces renormalization of the frequencies (Lamb-shifts) along with
the couplings between the oscillators. The resulting effective Hamiltonian can again be brought to the normal mode general form (\ref{NMbasis}) by a rotation as discussed above.

 Anticipating this  renormalization of the normal mode frequencies  and  mode coupling induced by the interactions with the bath, and following the arguments leading to the general form (\ref{NMbasis})  we write the   system Hamiltonian in the \emph{renormalized} normal mode basis corresponding to the \emph{renormalized} frequencies   by introducing a \emph{matrix of counterterms} $\delta \mathbf{\Omega}$  and write the total Hamiltonian as
 \be H_{tot} = \frac{p^2_+}{2}  + \frac{p^2_-}{2} + \frac{\Omega^2_{+R}}{2}~q^2_+ + \frac{\Omega^2_{-R}}{2}~q^2_- + (q_+~ q_-)~ \delta \mathbf{\Omega}~  \left(
                                        \begin{array}{c}
                                          q_+ \\
                                          q_- \\
                                        \end{array}
                                      \right) + H_B + H_{SB}\,,\label{renHS}
  \ee where $\Omega_{\pm R}$ are the \emph{renormalized} normal mode frequencies and
  \be \delta \mathbf{\Omega} =  \left(
                                  \begin{array}{cc}
                                    \delta \Omega_{++} & \delta \Omega_{+-} \\
                                    \delta \Omega_{+-} & \delta \Omega_{--} \\
                                  \end{array}
                                \right) \label{delct} \ee is a \emph{counterterm} frequency matrix that will be
 required to cancel the   contributions from the real part of the self-energy corrections (Lamb-shifts) that diverge in the limit of large bandwidths of the baths.

 Where $H_{SB}$ in eqn. (\ref{renHS}) is given by (\ref{SBNM}) in terms of  the \emph{renormalized} angle $\psi$.   Since the self-energies associated with the normal modes are given by (\ref{pluplu}-\ref{plumin}) and the counterterms are chosen to cancel the divergent contributions from these, we write following eqns. (\ref{pluplu}-\ref{plumin})
 \bea \delta \Omega_{++} & = &  \cos^2(\psi)\, \delta \Omega_{1}+ \sin^2(\psi)\,\delta \Omega_{2} \label{delpluplu}\\
 \delta \Omega_{--} & = &  \sin^2(\psi)\, \delta \Omega_{1}+ \cos^2(\psi)\,\delta \Omega_{2} \label{delminmin} \\
  \delta \Omega_{+-} & = &  \delta \Omega_{-+} = \cos(\psi)\, \sin(\psi) \Big[\delta \Omega_{1}-\delta \Omega_{2}\Big] \,, \label{delplumin} \eea where $\delta \Omega_{1,2}$ will be chosen to cancel the divergent contribution from $\Sigma_{1,2}$.
  This aspect will be discussed in detail below for the case of Ohmic baths (see section  \ref{sec:Ohmic}) but similar considerations should apply for any spectral density of a bath with a  bandwidth large compared to the (renormalized) frequencies in the system.

 Including this counterterm matrix in $H_S$, the matrix $\mathbf{G}^{-1}(s)$ (\ref{invG}) for the equations of motion in Laplace variable now reads
  \be   \mathbf{G}^{-1}(s) = \left(
                              \begin{array}{cc}
                                s^2+\Omega^2_{+R} + \delta \Omega_{++}+ \tilsigpp(s) & \delta \Omega_{+-}+\tilsigpm(s) \\
                                \delta \Omega_{-+}+\tilsigmp(s) & s^2+\Omega^2_{-R} + \delta \Omega_{--}+\tilsigmm(s) \\
                              \end{array}
                            \right) \,. \label{invGR}\ee The solution for $\tilde{q}_{\pm}(s)$ is obtained by inverting this matrix.
                                                      It is convenient to introduce the quantities
\be W^2_+(s) =   \Omega^2_{+R} + \delta \Omega_{++}+ \tilsigpp(s) ~~ ; ~~ W^2_-(s) =   \Omega^2_{-R} + \delta \Omega_{--}+\tilsigmm(s) \label{Wpm}\ee

\be M^2(s) = s^2 + \frac{1}{2}\Big[ W^2_+(s) + W^2_-(s)\Big] \,\label{M2}\ee
\be \Theta^2(s) = \delta \Omega_{+-} + \Sigma_{+-}(s) \, \label{teta2}\ee
\be \rho(s) = \Big[ \Big(  W^2_+(s) - W^2_-(s)\Big)^2 +  \Big(2\Theta^2(s)\Big)^2 \Big]^{1/2} \, \label{rho}\ee
\be \alpha(s) = \frac{\Big(W^2_-(s) - W^2_+(s)\Big)}{\rho(s)}~~;~~\beta(s) = \frac{2\Theta^2(s)}{\rho(s)}~~;~~ \alpha^2(s)+\beta^2(s)=1 \,, \label{alfabeta}\ee in terms of which one finds
\be \mathbf{G}^{-1}(s) =  M^2(s) \,\mathds{1} + \frac{\rho(s)}{2} ~ \left(
                                                                      \begin{array}{cc}
                                                                        -\alpha(s) & \beta(s) \\
                                                                        \beta(s) & \alpha(s) \\
                                                                      \end{array}
                                                                    \right)\,.  \label{gmin1oh}
 \ee

\noindent It is now straightforward to obtain
\be \mathbf{G}(s) = \frac{1}{2} \frac{\Big[\mathds{1}+\mathds{R} \Big]}{\Big[M^2(s)- \frac{\rho(s)}{2}\Big]}+\frac{1}{2} \frac{\Big[\mathds{1}-\mathds{R} \Big]}{\Big[M^2(s)+ \frac{\rho(s)}{2}\Big]}\,,
\label{gofs} \ee where $\mathds{1}$ is the identity matrix and
\be \mathds{R} = \left(
                   \begin{array}{cc}
                     \alpha(s) & -\beta(s) \\
                     -\beta(s) & -\alpha(s) \\
                   \end{array}
                 \right)\,. \label{mtxR}\ee  The matrix $\mathds{R}$ is traceless with  determinant $ (-1)$, hence its eigenvalues are $\lambda = \pm 1$. Therefore $(\mathds{1}\pm \mathds{R})/2$ are projectors onto the eigenvectors of $\mathds{R}$ with eigenvalues $\lambda = \pm 1$ respectively.

The solution of the Heisenberg-Langevin equations $q_\pm(t)$ can now be obtained by inverse Laplace transform, with
\be  \mathds{G}(t) = \int_{\mathcal{C}}\,\frac{ds}{2\pi i}~ e^{st}~\mathbf{G}(s) \label{goft} \,.\ee      The Bromwich contour $\mathcal{C}$ is parallel to the imaginary axis in the complex $s-$ plane to the right of all the singularities of $\mathbf{G}(s)$. Stability requires that the singularities have $\mathrm{Re}(s) \leq 0$; therefore   the Bromwich contour corresponds to $s = i\omega + \epsilon$ with $ \epsilon \rightarrow 0^+$ and
\be  \mathds{G}(t) = \int \frac{d\omega}{2\pi}~ e^{i\omega t}~\mathbf{G}(s=i\omega+\epsilon) \,. \label{gofteps} \ee

In order to obtain the Green's function matrix, we need the analytic continuation of the Laplace transform of the self-energies (\ref{sigmas}) to $s= i \omega + \epsilon$, with $\epsilon = 0^+$ understood. With (\ref{sigmaofs}) we find
\be     {\chi}_{1,2}(\omega)\, \equiv \, \widetilde{\Sigma}_{1,2}(s=i\omega+\epsilon)  = \frac{1}{\pi} \int_{-\infty}^{\infty} \, \frac{\sigma_{1,2}(\omega')}{\omega-\omega'-i\epsilon}\, {d\omega'} \,, \label{sigmaofome} \ee with
\be \mathrm{Im}\chi_{1,2}(\omega) = \sigma_{1,2}(\omega) ~~;~~ \mathrm{Re}\chi_{1,2}(\omega) = \frac{1}{\pi} \int_{-\infty}^{\infty} \,\mathcal{P}\,\Bigg[ \frac{\mathrm{Im}\,\chi_{1,2}(\omega')}{\omega-\omega' }\Bigg]\, {d\omega'} \,, \label{reimparts}\ee and $\chi_{\alpha \beta}$ are defined by the same linear combinations as $\Sigma_{\alpha \beta}$ for $\alpha, \beta = \pm$ given by eqns.(\ref{pluplu}-\ref{plumin}).

\noindent Therefore the solution for the time evolution of the normal mode Heisenberg operators is given by
\be \left(
      \begin{array}{c}
        q_+(t) \\
        q_-(t) \\
      \end{array}
    \right) = \mathds{G}(t) ~ \left(
      \begin{array}{c}
        \dot{q}_+(0) \\
        \dot{q}_-(0) \\
      \end{array}
    \right)~+ \dot{\mathds{G}}(t)~\left(
      \begin{array}{c}
        q_+(0) \\
        q_-(0) \\
      \end{array}
    \right)~ + \int^t_0 \mathds{G}(t-t')~\left(
                                           \begin{array}{c}
                                             \xi_+(t') \\
                                             \xi_-(t') \\
                                           \end{array}
                                         \right)~dt' \,. \label{qpmoft} \ee In order to highlight the separate contributions from initial conditions and noise terms, we write the solution (\ref{qpmoft}) as
 \be  \left(
      \begin{array}{c}
        q_+(t) \\
        q_-(t) \\
      \end{array}
    \right) =  \left(
      \begin{array}{c}
        q_+(t) \\
        q_-(t) \\
      \end{array}
    \right)_0 +  \left(
      \begin{array}{c}
        q_+(t) \\
        q_-(t) \\
      \end{array}
    \right)_{\xi} \label{split} \ee with
   \be    \left(
      \begin{array}{c}
        q_+(t) \\
        q_-(t) \\
      \end{array}
    \right)_0 = \mathds{G}(t) ~ \left(
      \begin{array}{c}
        \dot{q}_+(0) \\
        \dot{q}_-(0) \\
      \end{array}
    \right)~+ \dot{\mathds{G}}(t)~\left(
      \begin{array}{c}
        q_+(0) \\
        q_-(0) \\
      \end{array}
    \right) ~~;~~  \left(
      \begin{array}{c}
        q_+(t) \\
        q_-(t) \\
      \end{array}
    \right)_{\xi} =  \int^t_0 \mathds{G}(t-t')~\left(
                                           \begin{array}{c}
                                             \xi_+(t') \\
                                             \xi_-(t') \\
                                           \end{array}
                                         \right)~dt' \,. \label{splitininois}\ee
\noindent Correlation functions of Heisenberg operators $q_\alpha(t), \alpha = \pm$ require \emph{two} different averages:
\begin{itemize}
\item{   Average over the initial conditions denoted    by $\langle q_\alpha(t) q_\beta(t') \rangle_0$ correspond to averaging in the initial state in terms of the averages of the Heisenberg operators at the initial time $t=0$. We will \emph{assume} that the initial state is uncorrelated for the normal modes with $\langle q_\pm(0) \rangle_0 = \langle \dot{q}_\pm (0) \rangle_0 =0$ leading to $\langle q_\pm(t) \rangle_0 =0$. These assumptions on the initial state of the system can be relaxed with the corresponding expectation values and initial state correlations changing accordingly. }

\item{   Averages over the noise terms $\xi_\pm$ (see eqn. \ref{xis}) in terms of $\xi_1,\xi_2$, with the thermal noise averages given by eqn. (\ref{correnoise}). We will denote averages over the noise as $\langle \langle (\cdots) \rangle \rangle_\xi$. The bath averages (\ref{correnoise}) also  imply that $\langle \langle q_\pm(t) \rangle \rangle_\xi =0$. }

    Because the theory is Gaussian, only the one and two point correlation functions must be obtained; higher order correlation functions are obtained from Wick's theorem.
\end{itemize}

The results above describe how to extract the real time dynamics of relaxation and coherence in the general case of two mechanical oscillators coupled to each other as in  (\ref{Hsis}) and to respective thermal baths as in (\ref{SBcoup}).   For general spectral densities, the analysis of the time evolution, correlation functions and coherences will likely involve a numerical study. However, we can make analytic progress in the case of Ohmic baths described in section (\ref{sec:Ohmic}) below.

\section{Ohmic Baths}\label{sec:Ohmic}

We consider a Drude model for an Ohmic bath with
\be \sigma_{j}(\omega) = \mathrm{Im}\chi_{j}(\omega) = \omega \,  \gamma_{j} \,  \frac{\Lambda^2_{j}}{\Lambda^2_{j} + \omega^2} ~~;~~ j=1,2\label{drude} \ee from which we find
\be \mathrm{Re}\chi_{j}(\omega) = -\Lambda_{j} \,\gamma_{j} \,\frac{\Lambda^2_{j}}{\Lambda^2_{j} + \omega^2}~~;~~ j=1,2 \,. \label{realpartdrude}\ee Alternatively, we also consider the case of  sharp cutoffs for the respective spectral densities
\be \sigma_{j}(\omega) = \mathrm{Im}\chi_{j}(\omega) = \gamma_j \,\omega ~ \Theta(\Lambda_{j} - |\omega|) \,,\label{cutoff} \ee where $\Lambda_{j}$ are the cutoffs or bandwidths of the respective baths; in this case we find for the complex self-energy
\be \chi_{j}(\omega) = - \frac{2}{\pi}\,\gamma_{j}\,\Lambda_{j} - \frac{\omega}{\pi}\,\gamma_{j} \ln \Bigg|\frac{\Lambda_{j}-\omega}{\Lambda_{j}+\omega} \Bigg| + i \omega \, \gamma_{j}~~ ; ~~ \Lambda_{j} > |\omega| \,. \label{sigmacutoff} \ee In the limit $\Lambda_{j} \gg \omega$ both Ohmic spectral densities yield a  linearly ($\propto \Lambda_j$) divergent real part and the same imaginary part.  We are primarily interested in the low energy, long time dynamics which, after renormalization, should be insensitive to the  high frequency degrees of freedom of the bath. Therefore in the following we will consider the cut-off  spectral density   (\ref{cutoff}) for which the self-energies simplify, namely
\be \chi_{j}(\omega) = -\frac{2}{\pi}\,\gamma_{j}\Lambda_{j} + i\omega\,\gamma_{j}\,, \label{selfies}\ee where it is implicit that $\omega \ll \Lambda_j$. Following ref.\cite{weiss} we refer to this as the strict Ohmic case.

 We choose the counterterms $\delta \Omega_{j}$ to cancel the real part of the self-energies, namely
\be \delta \Omega_{j} + \mathrm{Re} \chi_{j}(\omega)  = 0  \Rightarrow \delta \Omega_{j} = \frac{2}{\pi}\, \gamma_{j}\Lambda_{j} \,,\label{renorma} \ee yielding in the normal mode basis
\be \delta \Omega_{++} + \mathrm{Re}\chi_{++}(\omega) = \delta \Omega_{--} + \mathrm{Re} \chi_{--}(\omega) = ~\delta \Omega_{+-} + \mathrm{Re} \chi_{+-}(\omega) =0 \,. \label{cancela}\ee

    It is convenient to define
\be \Omega_{R+} = W-\frac{\Delta}{2} ~~;~~\Omega_{R-} = W+\frac{\Delta}{2} \,,\label{detuning} \ee where $\Delta$ is the detuning between the normal modes.  For $\Delta =0$ the \emph{renormalized} normal modes are \emph{degenerate}, we emphasize that this   is different from the non-interacting case where the (un-renormalized) normal mode frequencies are given by eqn. (\ref{NMfreqs}) which become degenerate for $\Omega_a=\Omega_b=0; \Omega =0$.

Implementing the renormalization conditions (\ref{cancela}), with the definitions (\ref{detuning}) we find for ohmic baths described by the spectral densities  (\ref{cutoff})
\bea W^2_+(s)+W^2_-(s)  & = & 2 W^2 + \frac{\Delta^2}{2} + s\,(\gamma_1 + \gamma_2) \label{Wpluss}\\
W^2_-(s)- W^2_+(s)  & = & 2 W \Delta + s (\gamma_2 - \gamma_1) \,\cos(2\psi) \label{Wminss}\\
2\Theta^2(s) & = & -s (\gamma_2 - \gamma_1) \,\sin(2\psi)\,.\label{tetsqua}\eea It now remains to input these expressions into the Green's function to obtain the real time evolution; see below. Because there are several scales in the problem we focus on two limits of particular interest.

\vspace{2mm}

\subsection{Strong Coupling.}\label{sub:strongcoup}

We refer   to the case when the term $2W\Delta$ in (\ref{Wminss}) can be neglected as the strong coupling regime because the contributions from the couplings to the bath, which determine the off-diagonal terms  in the ``mixing matrix'' $\mathds{R}$ in (\ref{mtxR}) have the same magnitude as the diagonal terms. We study separately the cases $\Delta =0$ and $\Delta \neq 0 $ in the strong coupling regime because the case of vanishing detuning, $\Delta =0$,  is particularly relevant and describes a similar case in ref.\cite{keeling,marq}.

\subsubsection{$\Delta =0$}
\label{sub:delzero}


Vanishing detuning corresponds to degenerate \emph{renormalized} normal modes for the  mechanical oscillators. We note that because of renormalization effects from the interactions with the bath, the conditions of degeneracy of the renormalized normal modes
 are \emph{different} from that for degeneracy of the ``bare'' (unrenormalized) normal mode frequencies. From the relations  given by equations (\ref{delpluplu}-\ref{delplumin}) and (\ref{renorma}) one can find the relation between the bare frequencies and the spectrum of the baths that leads to the degeneracy of the renormalized normal modes.


 In this case, setting $\Delta = 0$ in eqns. (\ref{Wpluss}-\ref{tetsqua}) and with the definitions (\ref{rho}-\ref{mtxR})    we find
\be \mathbf{G}^{-1}(s) = \Big(s^2 + W^2 + \frac{s}{2}(\gamma_1+\gamma_2)\Big)~\mathds{1}+ \frac{s}{2}(\gamma_2-\gamma_1)\, \left(
                                                                                   \begin{array}{cc}
                                                                                     -\cos(2\psi) & -\sin(2\psi) \\
                                                                                     -\sin(2\psi) & \cos(2\psi) \\
                                                                                   \end{array}
                                                                                 \right) \,, \label{invGdelzero}\ee
from which we identify
\bea M^2(s) & = & \Big(s^2 + W^2 + \frac{s}{2}(\gamma_1+\gamma_2)\Big)~~;~~\rho(s)= s(\gamma_2-\gamma_1)\,, \nonumber \\
\alpha(s) & = & \cos(2\psi) ~~;~~ \beta(s) = -\sin(2\psi) \,,\label{paradelzero}\eea and
\be \mathds{R} = \left(
                   \begin{array}{cc}
                     \cos(2\psi) & \sin(2\psi) \\
                     \sin(2\psi) & -\cos(2\psi) \\
                   \end{array}
                 \right)  = \mathbf{V}(\psi) \, \left(
                                                       \begin{array}{cc}
                                                         1 & 0 \\
                                                         0 & -1 \\
                                                       \end{array}
                                                     \right)
                      \,\mathbf{V}^{-1}(\psi)\,,\label{Rdelzero} \ee where the matrix $\mathbf{V}(\lambda)$ is given by eqn. (\ref{Voflam}). Combining these results with the general expression for the Green's function matrix, equation (\ref{gofs}), we find
\be \mathbf{G}(s) =    \mathbf{V}(\psi) \, \left(
                                                       \begin{array}{cc}
                                                         g_1(s) & 0 \\
                                                         0 & g_2(s) \\
                                                       \end{array}
                                                     \right)
                      \,\mathbf{V}^{-1}(\psi) ~~;~~ g_{1,2}(s) = \frac{1}{\Big[s^2+  W^2 +s\,\gamma_{1,2}\Big]}\,, \label{gfindelzero} \ee which, upon using eqn. (\ref{gofteps}), yields
\be \mathds{G}(t) =    \mathbf{V}(\psi) \, \left(
                                                       \begin{array}{cc}
                                                         G_1(t) & 0 \\
                                                         0 & G_2(t) \\
                                                       \end{array}
                                                     \right)
                      \,\mathbf{V}^{-1}(\psi)  \equiv \frac{1}{2} \Big[G_1(t)+G_2(t) \Big] \mathds{1} +\frac{1}{2} \Big[G_1(t)-G_2(t) \Big]\mathds{R}  \label{Goftdelzero} \ee with
\be G_i(t) = e^{-\gamma_i t/2}~\frac{\sin\big[W_i\,t \big]}{W_i}~~;~~ W_i = \Big[W^2-\frac{\gamma^2_i}{4} \Big]^{1/2} ~~ i=1,2 \,. \label{Gis}\ee The solutions given by (\ref{qpmoft}), with the noise terms given by (\ref{xis}) and $\mathds{G}(t)$ given by (\ref{Goftdelzero}), become

\be \left(
      \begin{array}{c}
        q_+(t) \\
        q_-(t) \\
      \end{array}
    \right) = \mathds{G}(t) ~ \left(
      \begin{array}{c}
        \dot{q}_+(0) \\
        \dot{q}_-(0) \\
      \end{array}
    \right)~+ \dot{\mathds{G}}(t)~\left(
      \begin{array}{c}
        q_+(0) \\
        q_-(0) \\
      \end{array}
    \right)~ + \mathbf{V}(\psi)\,\int^t_0  ~\left(
                                           \begin{array}{c}
                                             G_1(t-t')\,\xi_1(t') \\
                                             G_2(t-t')\,\xi_2(t') \\
                                           \end{array}
                                         \right)~dt' \, .\label{qpmoftdelzero} \ee

\noindent The form of $\mathds{G}(t)$ eqn. (\ref{Goftdelzero}) indicates that performing the unitary transformation
\be \left(
      \begin{array}{c}
        q_1(t) \\
        q_2(t) \\
      \end{array}
    \right) = \mathbf{V}^{-1}(\psi)\, \left(
      \begin{array}{c}
        q_+(t) \\
        q_-(t) \\
      \end{array}
    \right) \,,\label{12basis} \ee   the solution in this new basis is given by
    \be q_i(t) = G_i(t) \,\dot{q}_i(0) + \dot{G}_i(t)\, q_i(0) + \int^t_0 G_i(t-t')\,\xi_i(t')\,dt' ~~;~~ i= 1,2 \,.\label{12solus}\ee


The interpretation of this result is clear: For vanishing detuning the normal modes are degenerate, therefore one can make a unitary transformation  (rotation) that diagonalizes the coupling to the independent baths $B_1, B_2$ in $H_{SB}$, eqn. (\ref{SBNM}). Equations (\ref{gfindelzero}) and (\ref{Goftdelzero})  clearly show  that the Green's function is diagonal in the $1,2$ \emph{bath basis}.   The normal mode coordinates $q_\pm(t)$ evolve as linear combinations of the $1,2$  modes,
which evolve
independently in time with simple complex frequencies.

We are now in position to obtain the correlation functions and coherences for the Ohmic case with zero detuning.

\noindent Assuming that the system is in the ground state $|0\rangle \equiv |0_+\rangle\,|0_-\rangle$  for the independent normal modes at $t=0$, and using $\dot{q}_\pm(0) = p_\pm(0)$ it follows that
\bea \langle 0|q_\pm(0)|0 \rangle &= & 0~~;~~\langle 0|  {p}_{\pm}(0) |0\rangle   =   0 \label{onepointcorr} \\
\langle 0| q^2_{\pm}(0) |0\rangle   & = &    \frac{1}{2\Omega_\pm} ~~;~~ \langle 0| q_{\pm}(0) q_{\mp}(0) |0\rangle = 0 \label{qscorr}\\
 \langle 0|  {p}^2_{\pm}(0) |0\rangle    &  =  &         \frac{\Omega_\pm}{2} ~~;~~ \langle 0| p_{\pm} (0) p_{\mp}(0) |0\rangle = 0 \label{pscorr}\\
 \langle 0|  {p}_{\pm}(0) q_{\pm}(0) |0\rangle   & = &  \frac{-i}{2} ~~;~~ \langle 0|  {p}_{\pm}(0) q_{\mp} |0\rangle =0 \,.\label{inicorrs} \eea  Because the initial conditions are given in terms of the normal mode basis, the representation of the solution given by (\ref{qpmoftdelzero}) is the most suitable to evaluate correlation functions. We are primarily interested in the two-point correlation functions of the normal mode variables as these yield information on coherence and thermalization. Since the statistical averages $\langle \langle \xi_{1,2} \rangle \rangle =0$, the two point correlators become a sum of the contribution from the average on the initial conditions and the statistical average of the noise. The solution (\ref{qpmoftdelzero}) is given  in terms of the contribution from initial conditions and noise as in eqns. (\ref{split}) with the initial condition term given in eqn. (\ref{splitininois}), and the noise term given by
 \be    \left(
      \begin{array}{c}
        q_+(t) \\
        q_-(t) \\
      \end{array}
    \right)_{\xi} =  \mathbf{V}(\psi)\,\int^t_0  ~\left(
                                           \begin{array}{c}
                                             G_1(t-t')\,\xi_1(t') \\
                                             G_2(t-t')\,\xi_2(t') \\
                                           \end{array}
                                         \right)~dt'  ~~ = \mathbf{V}(\psi)\,   \left(
      \begin{array}{c}
        q_1(t) \\
        q_2(t) \\
      \end{array}
    \right)_{\xi}\,. \label{ininoise}\ee

\begin{figure}[ht!]
\begin{center}
\includegraphics[height=4in,width=4in,keepaspectratio=true]{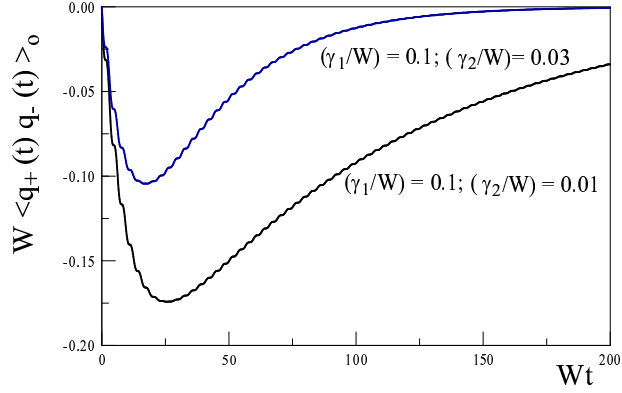}
\caption{ $W \,\langle q_+(t) q_-(t)\rangle_0   $  eqn. (\ref{iniave}) vs. $W\,t$ for $(\gamma_1/W)=0.1,(\gamma_2/W) = 0.01,0.03, \psi=\pi/4$.  }
\label{fig:qpqm}
\end{center}
\end{figure}

  Of particular importance is the \emph{coherence}\cite{keeling} or off-diagonal correlation function and a direct calculation using (\ref{splitininois}) along with  (\ref{onepointcorr}-\ref{inicorrs}) yields
\be \langle q_+(t)  q_-(t)  \rangle_0   = \frac{\sin(2\psi)}{4W}\Big[\big(\dot{G}^2_1(t)+W^2 G^2_1(t)\big)-\big(\dot{G}^2_2(t)+W^2 G^2_2(t)\big)  \Big] \,,\label{iniave} \ee where the subscript $0$ refers to expectation value in the initial (ground) state determined by (\ref{qscorr}-\ref{inicorrs}). Since $G_j(0)=0;\dot{G}_j(0)=1$, it is clear that $\langle q_+(0)q_-(0)\rangle_0 =0$ as determined by the initial conditions of independent normal modes (see second term in eqn. \ref{qscorr}). The behavior of $\langle q_+(t)  q_-(t)  \rangle_0 $ is shown in fig.(\ref{fig:qpqm}) for representative values of the parameters revealing transient coherence. The small scale oscillations revealed in the figure are at frequency $W$.

The noise contribution to the coherence is obtained from eqns. (\ref{ininoise},\ref{correnoise}) and is given by

\be \langle \langle q_+(t) q_-(t) \rangle \rangle_{\xi} = \frac{1}{2} \sin(2\psi) \int_0^t dt_1 \int^t_0 dt_2 \Big[ G_1(t-t_1)\,G_1(t-t_2) \langle \langle \xi_1(t_1)\,\xi_1(t_2) \rangle \rangle - G_2(t-t_1)\,G_2(t-t_2) \langle \langle \xi_2(t_1)\,\xi_2(t_2) \rangle \rangle\Big]\,. \label{diffchi}\ee

\noindent This expression manifestly highlights that the coherence vanishes if the baths have the same spectral densities ($\gamma_1 = \gamma_2$) \emph{and} temperature and also suggests its survival when the baths have \emph{different} spectral densities and/or  temperatures.

We now take the    cutoffs $\Lambda_j \rightarrow \infty$   and use the results (\ref{correnoise},\ref{invG} )  to  find for the noise contribution
\be \langle \langle q_+(t) q_-(t') \rangle \rangle_{\xi} = \sin(2\psi) \int^{\infty}_{-\infty} \frac{d\omega}{2\pi} \Big[I_1(\omega; t,t')\,\mathrm{Im}\chi_1(\omega)\,n_1(\omega) -
I_2(\omega; t,t')\, \mathrm{Im}\chi_2(\omega)\,n_2(\omega) \Big]\,, \label{noisecohe}\ee  where
\be I_j(\omega;t,t') = \int^t_0 dt_1 \int^{t'}_0 dt_2 \,G_j(t-t_1)\,G_j(t'-t_2)\,e^{i\omega(t_1-t_2)} ~~;~~ j=1,2\,.  \label{Isoft}\ee We consider a two-time correlation function to display the emergence of stationarity in the long time limit.  With $G_j(t)$ given by (\ref{Gis}) we find for vanishing detuning ($\Delta =0$) and in the asymptotically long time limit $t, t' \gg 1/\gamma_{1,2}$,
\be I_j(\omega;t,t') \rightarrow \frac{e^{i\omega(t-t')}}{\Big[\big(\omega^2-W^2 \big)^2+ \big(\omega\,\gamma_j \big)^2 \Big]}\,,  \label{asyIs} \ee  revealing the emergence of a stationary state at long times   under zero detuning conditions.   A similar calculation yields
\bea \langle \langle q_+(t) q_+(t') \rangle \rangle_{\xi} & = &  \int^{\infty}_{-\infty} \frac{d\omega}{ \pi} \Big[\cos^2(\psi)\,I_1(\omega; t,t')\,\mathrm{Im}\chi_1(\omega)\,n_1(\omega) + \sin^2(\psi)
I_2(\omega; t,t')\, \mathrm{Im}\chi_2(\omega)\,n_2(\omega) \Big] \label{qpp} \\
\langle \langle q_-(t) q_-(t') \rangle \rangle_{\xi} & = &  \int^{\infty}_{-\infty} \frac{d\omega}{ \pi} \Big[\sin^2(\psi)\,I_1(\omega; t,t')\,\mathrm{Im}\chi_1(\omega)\,n_1(\omega) + \cos^2(\psi)
I_2(\omega; t,t')\, \mathrm{Im}\chi_2(\omega)\,n_2(\omega) \Big]\,. \label{qmm} \eea

 Summarizing, in the stationary state, when  $t,t' \gg 1/\gamma_{1,2}$ the diagonal and off-diagonal  correlation functions  are given by
 \bea \langle \langle q_+(t+\tau) q_-(t) \rangle \rangle_{\xi} & = &  \sin( \psi) \,\cos(\psi) \Big[F_1(\tau)-F_2(\tau)] \label{etcoh} \\ \langle \langle q_+(t+\tau) q_+(t) \rangle \rangle_{\xi} & = & \cos^2(\psi) F_1(\tau) + \sin^2(\psi) F_2(\tau) \label{qppet}\\ \langle \langle q_-(t+\tau) q_-(t) \rangle \rangle_{\xi} & = & \sin^2(\psi) F_1(\tau) + \cos^2(\psi) F_2(\tau)  \,,\label{qmmet} \eea where
 \be F_j(\tau) = \int_{-\infty}^{\infty} \frac{d\omega}{\pi}  \, \frac{\mathrm{Im}\chi_j(\omega)\,n_j(\omega)\,e^{i\omega\tau} }{\Big[\big(\omega^2-W^2 \big)^2+ \big(\omega\,\gamma_j \big)^2 \Big]}~~;~~ j=1,2 \,. \label{Fdefs} \ee  For the Ohmic case (\ref{cutoff}) in the limit $\Lambda_j \rightarrow \infty$, one has  $\mathrm{Im}\chi_j(\omega) = \gamma_j \,\omega$  and    the integrals are finite. This  justifies  taking the infinite bandwidth limit first, since for $\Lambda_j \gg W, \gamma_j$   corrections are of   order   $(W/\Lambda_j)^2, (\gamma_j/\Lambda_j)^2$ and can be safely neglected. The detailed form of $F_j(\tau)$ is given in appendix (\ref{app:F}).

\noindent   In the asymptotic long time limit
$t,t'\gg 1/\gamma_{1,2}$  we see the emergence of a \emph{stationary   state} in the sense that the correlation functions become functions solely of the time difference.

 In the high temperature (classical) limit, $T_j \gg W, \gamma_j $,  and to leading order in the small ratios $W/T_j;\gamma_j/T_j \ll 1$,  we find the simple result
 \be F_j(\tau)  = \frac{T_j}{W^2}\,e^{-\gamma_j\tau/2}\,\Big[\cos[W_j\tau]+\frac{\gamma_j}{2W_j}\sin[W_j\tau]\Big] \,, \label{classFj}\ee  leading,  in the limit $\tau \rightarrow 0$, to the relations in the stationary state
 \bea \langle \langle q^2_+(t) \rangle \rangle_{\xi} & = &  \frac{1}{W^2}\Big[\cos^2(\psi)\,T_1 + \sin^2(\psi)\,T_2\Big] \label{qp2asy}\\
 \langle \langle q^2_-(t) \rangle \rangle_{\xi} & = &  \frac{1}{W^2}\Big[\sin^2(\psi)\,T_1 + \cos^2(\psi)\,T_2\Big] \label{qm2asy}\\
 \langle \langle q_+(t) q_-(t) \rangle \rangle_{\xi}  & = &    \frac{\sin(2 \psi)}{2 W^2} \, \Big[T_1 -T_2] \,. \label{coheclass}\eea In the above expressions we have neglected subleading terms suppressed by powers of  $W/T_j$ and $\gamma_j/T_j$ in the high temperature limit.

 \vspace{2mm}

 \textbf{Interpretation of results:} Although the result (\ref{coheclass}) is perhaps surprising and counterintuitive, the physical interpretation of equations (\ref{qp2asy}-\ref{coheclass}) is fairly simple: in terms of the coordinates $q_1,q_2$, namely the bath basis introduced by eqn. (\ref{12basis}),  the classical high temperature limit yields equipartition at temperatures $T_1,T_2$ respectively, in other words
 \be  \langle \langle q^2_1(t) \rangle \rangle_{\xi} = \frac{T_1}{W^2}~~;~~ \langle \langle q^2_2(t) \rangle \rangle_{\xi}= \frac{T_2}{W^2}\,. \label{equiparbathbasis}\ee The results (\ref{qp2asy}-\ref{coheclass}) follow from the relations $q_+ = \cos(\psi) \,q_1 -\sin(\psi) \,q_2~;~q_- = \sin(\psi) \, q_1 + \cos(\psi) \, q_2$,  and because $\xi_1,\xi_2$ are independent, so that  $\langle \langle q_1 q_2 \rangle \rangle_\xi =0 $ in the correlation functions of the normal mode coordinates (see eqn. (\ref{ininoise})). In particular the coherence
 \be \langle \langle q_+(t) q_-(t) \rangle \rangle_{\xi} = \frac{1}{2} \sin(2\psi) \big[\,  \langle \langle q^2_1(t) \rangle \rangle_{\xi}- \langle \langle q^2_2(t) \rangle \rangle_{\xi}\,\big] \,. \label{cohediffa}\ee

  In the present case $\Delta=0$ when the normal modes are degenerate,    the system Hamiltonian exhibits  a rotational symmetry in the $q_+,q_-$ plane and a rotation from the \emph{normal mode} basis to the \emph{bath basis} makes the \emph{total} Hamiltonian \emph{diagonal} in the bath basis $q_1,q_2$.  In this basis the asymptotic correlation functions describe a steady state in equilibrium with each corresponding bath with classical equipartition at high temperature. However,  the normal modes are linear combinations of the coordinates $q_1,q_2$ in the bath basis, hence the  correlation functions of the normal modes display the ``mixing'' of the correlation functions of $q_{1,2}$  in the bath basis.

 Whereas the coordinates in the bath basis $q_{1,2}$ reach a stationary state with high temperature equipartition at the respective temperatures $T_1,T_2$ of the   baths, the averages $\langle \langle q^2_{\pm}\rangle \rangle$ may be associated with \emph{effective} equipartition temperatures
 \be T^{eff}_+ \equiv \cos^2(\psi) T_1 + \sin^2(\psi) T_2 ~~;~~ T^{eff}_- \equiv \sin^2(\psi) T_1 + \cos^2(\psi) T_2 \,. \label{Teffs}\ee Furthermore, this analysis clearly indicates that the coherence $ \langle \langle q_+(t) q_- (t) \rangle \rangle_\xi$ is a result of the off diagonal couplings of the mechanical normal modes to the baths. Since  the mechanical normal modes reach a stationary state with high temperature equipartition at different (effective) temperatures, we refer to this   as a \emph{non-equilibrium stationary state}.

We refer to the stationary state   when both baths are at the same temperature, $T_1=T_2 \equiv T$, as the \emph{equilibrium case}, wherein  the averages (\ref{qp2asy},\ref{qm2asy}) agree with classical equipartition at high temperature $T\gg W$, and the coherence (\ref{coheclass}) vanishes to leading order in $W/T_j~,~\gamma_j/T_j$.

 We conclude that for vanishing detuning and in the   high temperature (classical) limit of both baths in the \emph{non-equilibrium} case with  $T_1 \neq T_2$, the coherence between the mechanical normal modes induced by the coupling to
the baths does not vanish in the long time limit $t \gg 1/\gamma_{1,2}$, and is large for $|T_1-T_2|\gg W$.  Therefore, the survival of coherence between the mechanical normal modes in the high temperature limit of both baths is a consequence of the \emph{non-equilibrium} nature of the asymptotic stationary state for $T_1 \neq T_2$.

 A similar analysis for the   correlation functions and coherence of the canonical momenta $p_\pm$ yields for the noise averages in the stationary regime
 \bea \langle \langle p_+(t+\tau) p_-(t) \rangle \rangle_{\xi} & = &  \sin( \psi) \,\cos(\psi) \Big[H_1(\tau) -H_2(\tau)] \label{etcohppm} \\ \langle \langle p_+(t+\tau) p_+(t) \rangle \rangle_{\xi} & = & \cos^2(\psi) H_1(\tau) + \sin^2(\psi) H_2(\tau) \label{etppp}\\ \langle \langle p_-(t+\tau) p_-(t) \rangle \rangle_{\xi} & = & \sin^2(\psi) H_1(\tau) + \cos^2(\psi) H_2(\tau)  \,,\label{etpmm} \eea where
 \be H_j(\tau) = \int_{-\infty}^{\infty} \frac{d\omega}{\pi}  \, \frac{ \omega^2\,\mathrm{Im}\chi_j(\omega)\,e^{i\omega\tau} n_j(\omega)  }{\Big[\big(\omega^2-W^2 \big)^2+ \big(\omega\,\gamma_j \big)^2 \Big]} = -\frac{d^2}{d\tau^2}\, F_j(\tau) ~~;~~ j=1,2 \,. \label{Hdefs} \ee

 For the strictly Ohmic case with $\mathrm{Im}\chi_j(\omega) = \gamma_j \,\omega$,  as $\tau \rightarrow 0$ the extra factor $\omega^2$ in the  frequency integral yields a  logarithmic  divergence with  the upper cutoffs $\Lambda_j$. This divergence can be best seen in the $T_j \rightarrow 0$ limit where $n_j(\omega) \rightarrow -\Theta(-\omega)$ (with $\Theta$ the step function) by cutting off the integrals at $\Lambda_j\gg W,\gamma_j$. In this limit we find

  \be H_j(0) =  \frac{\gamma_j}{\pi}\,\ln\Big(\frac{\Lambda_j}{W}\Big)-\frac{W}{2\pi\sqrt{\frac{\gamma^2_j}{W^2}-4}}
  \Bigg[Z_+\,\ln[-Z_+] - Z_-\,\ln[-Z_-] \Bigg]~~;~~ Z_\pm = \frac{1}{2}\Bigg[2-\frac{\gamma^2_j}{W^2} \pm \frac{\gamma_j}{W}\sqrt{\frac{\gamma^2_j}{W^2}-4}~ \Bigg] \,.   \label{Hjcutoff}\ee  If instead we use the Drude form (\ref{drude})  we find the same result  for $\Lambda_j \gg W,\gamma_j$, namely a logarithmic dependence on the bandwidth of the bath. The logarithmic dependence on $\Lambda_j$ is also a feature if   an exponential cutoff is used, of the form   $\mathrm{Im}\chi_j(\omega) = \gamma_j \,\omega\,e^{-|\omega|/\Lambda_j}$, as can be easily verified in the limit $\Lambda_j \gg W, \gamma_j$. Therefore, this divergence is \emph{not} an artifact of a sharp cutoff but a general result of the Ohmic case when the bandwidth of the bath is much larger than the typical scales (relaxation and oscillations) of the system. This results in a logarithmic dependence on the bandwidth  for the expectation value of the energy of the normal modes. This divergence with the bath bandwidth has been recognized in ref.\cite{weiss,quthermo1}, and for free Brownian motion in \cite{breuer}, and more recently in \cite{caldeirarecent}. In the \emph{sub-ohmic } case when $\mathrm{Im}\chi_j(\omega) \propto \omega^s,~~0< s < 1$ the integral is no longer divergent, but correlation functions with  higher order $\tau$ derivatives eventually will develop a divergent bandwidth dependence. The super-Ohmic case $s>1$ will feature stronger dependence on the bandwidth.

 At finite temperature and $\tau \neq 0$ we find
 \be H_j(\tau) = H^{(a)}_j(\tau) + H^{(b)}_j(\tau) \,,\label{H1H2}\ee where
 \be H^{(a)}_j(\tau)  = -\frac{d^2}{d\tau^2}\, F^{(a)}_j(\tau)\,, \label{Haoftau} \ee and  $F^{(a)}_j(\tau)$ is given by (\ref{F1js}) and is obviously independent of the cutoff. Similarly, we find
 \be H^{(b)}_j (\tau) = \frac{\gamma_j}{ \pi}~\sum_{l=1}^{\infty} \frac{e^{-2\pi l  T_j \tau}}{l \,\Bigg[ \Big(1+\big(\frac{W}{2\pi l T_j}\big)^2 \Big)^2 - \Big( \frac{\gamma_j}{2\pi l T_j}\Big)^2 \Bigg]}\,. \label{Hboftau}\ee This last contribution is subdominant for $T_j \gg \gamma_j$.
Clearly \emph{each term} in the sum over $l$ in (\ref{Hboftau}) is finite in the $\tau \rightarrow 0$ limit; however, the nature of the divergence in this limit   is gleaned from the asymptotic $1/l$ behavior  behavior of the series  for large $l$  recognizing that
\be   \sum_{l=1}^\infty \frac{e^{-2\pi l  T_j \tau}}{l} = -\ln[1-e^{-2\pi T_j \tau}] ~~~ \overrightarrow{\tau \rightarrow 0 }~~~ \ln[\frac{1}{2\pi T_j \tau}]\,. \label{logdiv}\ee

Physically, however,   time scales of order $\tau \lesssim 1/\Lambda_j$ probe  the high frequency degrees of freedom of the bath;  taking as a minimum time scale $\tau \simeq 1/\Lambda_j$ the result (\ref{logdiv}) yields
\be H^{(b)}_j (\tau \simeq \Lambda^{-1}_j) = \frac{\gamma_j}{\pi}\,\ln\Big[\frac{\Lambda_j}{2\pi T_j}\Big] + \cdots \label{asyHb} \ee where the dots stand for terms that remain finite in the $\Lambda_j \rightarrow \infty $ limit. Hence, for $\tau \lesssim 1/\Lambda_j$   the leading logarithmic dependence on the bandwidth of the baths is precisely the one obtained in the zero temperature limit (\ref{Hjcutoff}).

 The time evolution of correlation functions in the stationary (but non-equilibrium) regime becomes insensitive to the bandwidth and high frequency degrees of freedom of the bath for time scales $\tau \gg 1/\Lambda_j$. In the limit $\Lambda_j \gg W_j, \gamma_j, \Omega_\pm$, namely   exceeding the corresponding frequency or energy scales of the system, there is a wide separation of time scales.  For $\Lambda_j \tau \gg 1$ and $W_j\,\tau~;~\gamma_j \tau \simeq 1$ the correlations do not reflect the high frequency modes of the bath, and one can construct a universal (in that sense) effective low frequency/energy  theory for
the system after ``integrating out'' the bath degrees of freedom.

 The logarithmic dependence of the momentum correlation function in the short time (coincidence) limit in the Ohmic case has recently been been exploited to yield a bound on the ``logarithmic negativity'' as a measure of entanglement\cite{marq}.

 From these results we conclude that $\tau \neq 0$ effectively regulates the high frequency contribution from the spectral density of the baths. Physically for $t\gg 1/\gamma_{1,2}$ as $\tau \rightarrow 0$ these correlations probe  the dynamics of the high frequency  degrees of freedom of the bath in the stationary state. In order to obtain an effective low energy and long time description of the system (after ``integrating out the bath''),  a coincidence limit of these correlations  in the stationary regime ($t \gg 1/\gamma_{1,2}~;~ \tau \rightarrow 0$)  must be interpreted as $\tau \,\Lambda_j \gg 1$ and $\tau \,\Omega_\pm$ , $\tau \,\gamma_{1,2} \ll 1$.

\subsubsection{One bath case, $\Delta =0$ }\label{subsec:onebath}

The case of the oscillators coupled to only one bath, say $B_1[\{Q_1\}]$ considered in ref.\cite{keeling} is obtained by simply setting $B_2[\{Q_1\}]=0$. In other words one sets $\mathrm{Im}\chi_2(\omega)=0$ and $ \xi_2 =0$ in the above results which, for the Ohmic case translate into $\gamma_2=0$, yielding, in the case $\Delta =0$, one damped mode and one  \emph{undamped mode} with
\be G_1(t) = e^{-\gamma_1 t/2}~\frac{\sin\big[W_1\,t \big]}{W_1}~~;~~G_2(t) =   \frac{\sin\big[W\,t \big]}{W}\,. \label{g21bath}  \ee  in the Green's function (\ref{Goftdelzero}).  The undamped mode is expected because of the degeneracy of the mechanical oscillators in the case of vanishing detuning. A unitary transformation to the bath basis $1,2$ in which $H_S$ remains diagonal for $\Delta=0$,  immediately leads to the conclusion that the mode $2$ is undamped.

 This conclusion also follows directly from the Green's functions (\ref{gfindelzero},\ref{Goftdelzero}): setting $\gamma_2=0$ the diagonal component $g_2(s)$ features an undamped pole, this is also manifest in $G_2(t)$ which is the Green's function for an undamped oscillator. The normal mode coordinates $q_\pm(t)$ evolve in time as linear combinations of the bath basis modes $1,2$, one damped and one undamped.

 In the asymptotic long time limit $\gamma_1 t \gg 1$ inserting (\ref{g21bath})  into (\ref{iniave}) we find
\be  \langle q_+(t) q_-(t) \rangle_0 \rightarrow  -\frac{\sin(2\psi)}{4W} \,. \label{iniave1bath} \ee

\begin{figure}[ht!]
\begin{center}
\includegraphics[height=4in,width=4in,keepaspectratio=true]{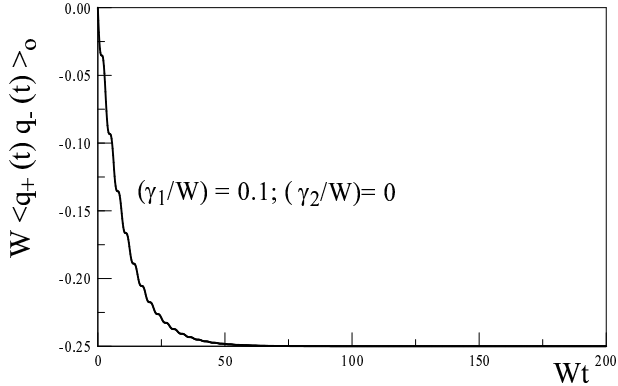}
\caption{ $W \,\langle q_+(t)  q_- (t)\rangle_0   $  eqn. (\ref{iniave}) vs. $W\,t$ for $(\gamma_1/W)=0.1,(\gamma_2/W) = 0, \psi=\pi/4$.  }
\label{fig:gqpqmgama2zero}
\end{center}
\end{figure}

This equal time correlation function is displayed in fig. (\ref{fig:gqpqmgama2zero}) for $\gamma_2=0$ and should be compared to figure (\ref{fig:qpqm}). For the noise contributions we find,
\be \langle \langle q_+(t) q_-(t') \rangle \rangle_{\xi} =  {\sin( \psi)\cos(\psi)}  \,F_1(t-t') ~~;~~ \langle \langle q_+(t) q_+(t') \rangle \rangle_{\xi} = \cos^2(\psi)\,F_1(t-t')~~;~~\langle \langle q_-(t) q_-(t') \rangle \rangle_{\xi} = \sin^2(\psi)\,F_1(t-t') \label{qq1bath}\ee where
\be F_1(t-t') = \int^{\infty}_{-\infty} \frac{d\omega}{ \pi} \frac{ \mathrm{Im}\chi_1(\omega)\,n_1(\omega)}{\Big[\big(\omega^2-W^2 \big)^2+ \big(\omega\,\gamma_1 \big)^2 \Big]} ~~e^{i\omega(t-t')}    \,, \label{noisecohe2}\ee which yields the results given by eqns. (\ref{F1js}, \ref{F2js}) with $j=1$.

Our results (\ref{qq1bath},\ref{noisecohe}) differ from those in ref.\cite{keeling} in that we \emph{do not find} a singularity in the asymptotic correlation functions and coherence.  We trace the origin of this discrepancy to the expressions (\ref{noisecohe},\ref{qpp},\ref{qmm}) wherein we see that all the correlation functions require the product (\ref{Isoft}) which is \emph{diagonal} in the product of Green's functions; we do not find an  interference (cross term) of the form $G_1 G_2$ which could yield such singularity. Also when $B_2[\{Q_2\}]=0$ the full contribution from the self energy $2$ vanishes identically.

An important corollary in this case with one undamped mode is that the system never loses information on the initial conditions. Whereas the asymptotic equal time limit of the noise contribution is completely determined by the spectral properties of the bath, had we chosen initial conditions different from those leading to (\ref{onepointcorr}-\ref{inicorrs}) the asymptotic limit of the initial state contribution (\ref{iniave1bath}) would reflect the different choice. This asymptotic dependence on the initial conditions, an obvious consequence of the undamped mode, suggests that a Markovian approximation to the time evolution of the reduced density matrix (neglecting the ``history'' of the dynamical evolution) \emph{may} break down in this particular case\cite{keeling}.

\subsubsection{   Non-vanishing detuning, $\Delta \ne 0$  }\label{subsec:2bathdet}

For non-vanishing detuning $\Delta \neq 0$ the degeneracy between the normal modes of the mechanical oscillators is lifted. We have carried out a perturbative expansion for $\Delta \neq 0$ in the ratio $ \Delta/|\gamma_2-\gamma_1|$. The result of this exercise is   cumbersome and not very illuminating, with the only noteworthy feature that in the \emph{one-bath  case} the mode that is undamped for $\Delta =0$ acquires a small damping rate $\propto \Delta^2/(|\gamma_2-\gamma_1|)$, and the initial condition contribution to the coherence is given by (\ref{iniave}) now with $G_2(t)$ exponentially damped with a small     damping rate $\propto \Delta^2/(|\gamma_2-\gamma_1|)$. Therefore, for $\Delta \neq 0$, it follows that
 $\langle q_+(t) q_-(t)\rangle_0 \rightarrow 0$ in the asymptotic long time limit consistent with the loss of the undamped mode. However, figures (\ref{fig:qpqm}) and (\ref{fig:gqpqmgama2zero})) indicate  that in the limit of non-vanishing but small detuning  coherence will survive for a long time determined by the longest time scale $\propto |\gamma_2-\gamma_1|/\Delta$ .  Furthermore,   as $\Delta \rightarrow 0$ we find a  smooth limit to the results presented above in the stationary regime. Therefore, whereas we do not find any singularity in the limit $\Delta \rightarrow 0$, our results are in broad agreement with those found in ref.\cite{keeling} in that the vanishing detuning case with both mechanical oscillators coupling to only one bath features an undamped mode, and that this peculiar feature    \emph{may} signal the breakdown of a Markovian approximation for the quantum master equation of the reduced density matrix. In the formulation followed here, we find that the  undamped mode leads to correlation functions that retain memory of the initial conditions; this is particularly clear in the case of the coherence which is given by eqn. (\ref{iniave1bath}) for the initial conditions given by (\ref{onepointcorr}-\ref{inicorrs}). Other initial conditions will yield different results for the asymptotic correlations and coherence. A small perturbation away from $\Delta=0$ yields a small relaxation rate for the mode that is undamped in the case of vanishing detuning, and asymptotically at long time (much longer than the inverse of the small relaxation rate) the correlations and coherence loose memory of the initial conditions. The noise contribution to  the averages (\ref{qp2asy},\ref{qm2asy}) and coherence (\ref{coheclass}) are similar to that of the $\Delta =0$ case with the addition of small perturbative corrections proportional to $\Delta^2/(\gamma_2-\gamma_1)^2$. However, a noteworthy feature is that these   perturbative corrections yield a  small but non-vanishing coherence in the equilibrium case $T_1=T_2\equiv T$ in the high temperature regime $T \gg W,\Delta$. Therefore, for $\Delta \neq 0$ memory of the initial condition remains for a long time and asymptotically the coherence does not vanish  in the equilibrium case but is  strongly suppressed by factors $\Delta^2/(\gamma_2-\gamma_1)^2 \ll 1$. Another example of such perturbative aspect is discussed below.

\subsection{Weak coupling}\label{subsec:weakcoup}

We now consider the weak coupling regime with $W, |\Delta|  \gg \gamma_{1,2}$ and expand to leading order in the ratios $\gamma_j/ {|\Delta|}; \gamma_j/W ~;~j=1,2$.  This case provides the groundwork for a future study of the master equation in the weak damping limit. In terms of the normal mode average frequency and detuning $W, \Delta$, introduced in eqn. (\ref{detuning}) we find
\bea \rho(s) & \simeq &  2W \Delta + s (\gamma_2-\gamma_1) \,\cos(2\psi) \label{rhowc}\\
\alpha(s) & \simeq & 1-\frac{1}{2}\Big(\frac{s}{2 W \Delta}\Big)^2\,(\gamma_2-\gamma_1)^2\,\sin^2(2\psi) \label{alfawc}\\
\beta(s) & \simeq &  -\frac{s}{2W\Delta}\,(\gamma_2-\gamma_1) \,\sin(2\psi) \,,\label{betawc} \eea and

\bea M^2(s)-\frac{\rho(s)}{2} & = & s^2 +\Omega^2_{+ } + s \Gamma_+ \label{mrhopwc}\\
M^2(s)+\frac{\rho(s)}{2} & = & s^2 +\Omega^2_{- } + s \Gamma_- \,,\label{mrhomwc} \eea where $\Omega_\pm$ are the (renormalized) frequencies of the normal modes of the mechanical oscillators, and
\be \Gamma_+ = \gamma_1 \cos^2(\psi) + \gamma_2 \sin^2(\psi)~~;~~ \Gamma_- = \gamma_2 \cos^2(\psi) + \gamma_1 \sin^2(\psi) \,.\label{Gamapm}\ee

\begin{figure}[ht!]
\begin{center}
\includegraphics[height=4in,width=4.5in,keepaspectratio=true]{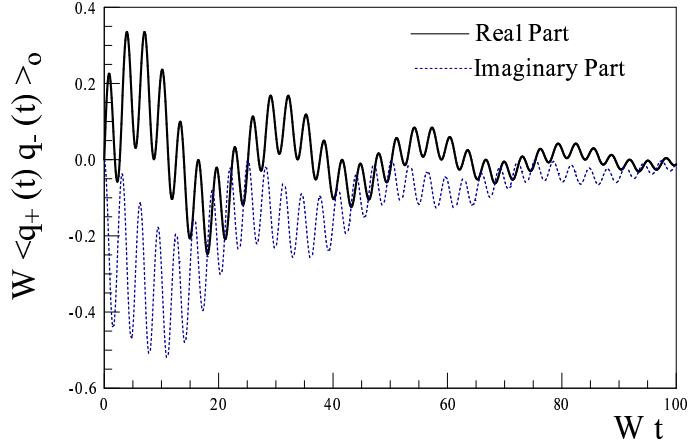}
\caption{ $W \,\langle q_+(t)  q_- (t)\rangle_0  $  eqn. (\ref{qoffdiagWC}) vs. $W\,t$ for $(\gamma_1/W)=0.05,(\gamma_2/W) = 0.005, \psi=\pi/4,\Delta/W = 0.25$.  }
\label{fig:beats}
\end{center}
\end{figure}

\noindent To leading order in $\gamma_j/ |\Delta|,\gamma_j/W$ we find
\be \mathds{R}(s) \simeq \left(
                           \begin{array}{cc}
                             1 & -\beta(s) \\
                             -\beta(s) & -1 \\
                           \end{array}
                         \right)~~;~~ \beta(s) \simeq -\frac{s}{2W\Delta}\,(\gamma_2-\gamma_1) \,\sin(2\psi)\,, \label{RWC}\ee  where we have neglected terms of order $(s(\gamma_2-\gamma_1)/2W\Delta)^2$ in $\alpha(s)$. A straightforward calculation yields to leading order
  \be \mathds{G} (t) = \left(
                         \begin{array}{cc}
                           g_+(t) & 0 \\
                           0 & g_-(t) \\
                         \end{array}
                       \right)+\frac{(\gamma_2-\gamma_1)}{2 W \Delta}\,\sin(2\psi)\, h(t) \, \left(
                                                   \begin{array}{cc}
                                                     0 & 1 \\
                                                     1 & 0 \\
                                                   \end{array}
                                                 \right)\,, \label{GoftWC}\ee where
 \be g_{\pm}(t) = e^{-\Gamma_\pm t/2} ~\frac{\sin(\Omega_\pm t)}{\Omega_\pm}\,, \label{litgoftWC}\ee and to leading order in  $\Gamma_\pm/\Omega_\pm$ we find

\be   h(t) = e^{-\Gamma_+ t/2}\,\cos(\Omega_{+ } t) - e^{-\Gamma_- t/2}\,\cos(\Omega_{- } t) \,. \label{delht}\ee  To leading order, the contribution to the off - diagonal correlator (coherence) from initial conditions is  found to be
\bea \langle q_+(t)q_-(t) \rangle_0 = && i \,\frac{(\gamma_2-\gamma_1)}{2W\Delta}\,\sin(2\psi)\, \Bigg\{ e^{-\Gamma_+ t} + e^{-\Gamma_- t} - e^{-(\Gamma_++\Gamma_-)t/2}\,\Big[e^{-i\Omega_+t}\big(\cos(\Omega_- t)+ i \frac{\Omega_-}{\Omega_+}\sin(\Omega_- t)\big) \nonumber \\ + &&  e^{i\Omega_- t}\big(\cos(\Omega_+ t)+  i \frac{\Omega_+}{\Omega_-}\sin(\Omega_+ t)\big)\Big]  \Bigg\} \,. \label{qoffdiagWC}\eea This correlation function displays interference beats between the normal modes reflecting a coherence that emerges from  the mixing of the normal modes mediated by their  coupling to the baths. An example is displayed in fig. (\ref{fig:beats}) for a range of parameters consistent with the weak coupling limit.

   Although the contributions from the \emph{initial conditions} do not lead to coherence in the asymptotic long time limit, in the weak coupling limit for  $\Gamma_+ + \Gamma_- = \gamma_1 + \gamma_2 \ll  {|\Delta|} $, it survives for a long time and displays interference between the (renormalized) normal modes. The interference beats may be observable during several periods before being damped out.

Since the spectral densities are proportional to $\gamma_{j}$, the leading order contribution to the correlation functions induced by the noise $\langle \langle (\cdots) \rangle \rangle_\xi$ is obtained by keeping only the diagonal term in $\mathds{G}(t)$ in (\ref{GoftWC}), leading to the result
\be \left(\begin{array}{c}
      q_+(t) \\
      q_-(t)
    \end{array}\right)_\xi = \left(\begin{array}{c}
      \int_0^t g_+(t-t') \xi_+(t') dt' \\
      \int_0^t g_-(t-t') \xi_-(t') dt'
    \end{array}\right)\,, \label{qsWCnoise} \ee where $\xi_{\pm}$ are defined by eqn. (\ref{xis}).

   Now, using equations (\ref{xis}, \ref{correnoise}, \ref{cutoff}) and (\ref{GoftWC}), and taking the cutoffs $\Lambda_j \rightarrow \infty$,  we now find for $t \gg 1/\gamma_{1,2}$ the emergence of a stationary regime for which
    \bea \langle \langle q_+(t+\tau) q_+(t ) \rangle \rangle_\xi & = &  \cos^2(\psi)\,J^{(1)}_{++}(\tau)+ \sin^2(\psi)\,J^{(2)}_{++}(\tau) \label{qppWC} \\
      \langle \langle q_-(t+\tau) q_-(t) \rangle \rangle_\xi  & = &    \sin^2(\psi)\,J^{(1)}_{--}(\tau)+ \cos^2(\psi)\,J^{(2)}_{--}(\tau) \label{qmmWC} \\
 \langle \langle q_+(t+\tau) q_-(t) \rangle \rangle_\xi & = & \cos(\psi)\sin(\psi)\,\Big[J^{(1)}_{+-}(\tau)- J^{(2)}_{+-}(\tau)\Big] \, \label{qpmWC} \eea  where the functions $J^{(j)}_{\alpha \beta}(\tau)\, , \alpha, \beta = \pm$ are given in Appendix (\ref{app:J}). Eqn. (\ref{qpmWC}) shows that the coherence depends on the difference of the spectral function and occupation of the baths and vanishes if the baths feature the same spectral function and temperature.

 Using the results in this appendix, we find that in the asymptotic stationary state, for $t \gg 1/\gamma_{1,2}$, the correlation functions of the momenta are related to those of the coordinates by
 \be  \langle \langle p_\alpha(t+\tau) p_\beta(t ) \rangle \rangle_\xi = -\frac{d^2}{d\tau^2} \Big[ \langle \langle q_\alpha(t+\tau) q_\beta(t ) \rangle \rangle_\xi \Big] ~~;~~ \alpha,\beta = +, - \,. \label{momcorres}\ee In particular for $\tau \lesssim 1/\Lambda_j$ these correlation functions have the same logarithmic dependence on the bandwidth of the baths as in (\ref{asyHb}), which as discussed in the previous subsection, is a manifestation of the same logarithmic dependence on cutoff found at zero temperature, (see eq. (\ref{Hjcutoff})). A finite $\tau$ regulates the high frequency behavior of the integrals, and  physically, as discussed above, a coincidence limit must be interpreted as the limit $\Lambda_j \tau \rightarrow \infty ~~;~~ \gamma_j \, \tau ~,~ \Omega_\pm \,\tau \rightarrow 0$.

     In the    limit  $\gamma_{1,2} \, t \rightarrow \infty $ but with $ \Lambda_{1,2}\, \tau \rightarrow \infty~;~   \Omega_\pm \tau \rightarrow 0$,  we find  to leading order in $\gamma_{1,2} \ll \Omega_{\pm}$ and in the high temperature   limit $T_{1,2} \gg \Omega_\pm, \gamma_{1,2}$,   the following results (see   Appendix (\ref{app:J}) for details)

    \bea \langle \langle q_+(t+\tau) q_+(t ) \rangle \rangle_\xi  & \simeq & \frac{1}{\Omega^2_+ \, \Gamma_+}\,\Big[\cos^2(\psi)\,T_1 \,\gamma_1+\sin^2(\psi)\,T_2 \,\gamma_2  \Big] \label{hitqppWC}\\\langle \langle q_-(t+\tau) q_-(t ) \rangle \rangle_\xi  & \simeq & \frac{1}{\Omega^2_-\, \Gamma_-}\,\Big[\sin^2(\psi)\,T_1 \,\gamma_1+\cos^2(\psi)\,T_2 \,\gamma_2  \Big] \label{hitqmmWC}\\\langle \langle q_+(t+\tau) q_-(t ) \rangle \rangle_\xi  & \simeq & \cos(\psi)\sin( \psi)  \,\Big[ \frac{\gamma_1+\gamma_2}{2(W\Delta)^2}\Big]\,\Big[  T_1 \,\gamma_1  -T_2 \,\gamma_2  \Big]  \,. \label{hitqpmWC}\eea The corrections to the above results in the high temperature limit are obtained from the complex poles of   $\coth[\omega/2T]$, and it is straightforward to find that these contributions are suppressed by the ratios    $\propto (\Omega_\pm/T_j)^2;\,(\gamma_j/T_j)^2$ with respect to the leading order terms (\ref{hitqppWC}-\ref{hitqpmWC}) (for example compare the terms $J^{(j,a)}_{\alpha \beta}~;~J^{(j,b)}_{\alpha \beta}$ in Appendix (\ref{app:J})).

    Using the relations (\ref{Gamapm}) one finds that in the \emph{equilibrium} high temperature case $T_1=T_2 \equiv T \gg W,\Delta$ and in the coincidence limit $\tau \rightarrow 0$, the averages (\ref{hitqppWC},\ref{hitqmmWC}) agree with classical equipartition, whereas the coherence becomes
    \be \langle \langle q_+(t) q_-(t) \rangle \rangle_\xi    \simeq   \cos(\psi)\sin( \psi)  \, \frac{T}{W^2}~ \Big[ \frac{\gamma^2_1-\gamma^2_2}{2 \Delta^2}\Big]\,   \,. \label{hitcohWC} \ee Although the coherence does not vanish in the high temperature classical regime in equilibrium and is enhanced by the factor $T/W$, it is strongly suppressed by the perturbative factor $(\gamma^2_1-\gamma^2_2)/\Delta^2 \ll 1$, but indicates a finite coherence in the classical limit in general.

\vspace{2mm}

\textbf{Interpretation of results:} The result (\ref{hitqpmWC}) also seems surprising and counterintuitive, yet its main feature, that the coherence survives in the high temperature limit for \emph{different baths}  also has a simple interpretation that stems from the perturbative relations (\ref{qsWCnoise}). From these it follows that
\be \langle \langle q_+(t) q_-(t) \rangle \rangle_\xi    \propto \langle \langle \xi_+ \xi_- \rangle \rangle   \label{offdi}\ee where $\xi_{\pm}$ are the linear combinations defined by eqn. (\ref{xis}). Therefore
\be \langle \langle \xi_+ \xi_- \rangle \rangle  = \cos(\psi)\sin(\psi) \Big[ \langle \langle \xi_1 \xi_1 \rangle \rangle - \langle \langle \xi_2 \xi_2 \rangle \rangle \Big] \,,  \label{bathnoi}\ee hence,  the coherence must vanish if both baths are identical, namely if they have the same spectral densities  \emph{and} temperatures $T_1=T_2$.

    \section{Bath mediated entanglement}\label{sec:entanglement}

    Harnessing quantum mechanical entanglement for quantum information and quantum computing is one of the primary foci of current experimental efforts, and as platforms that exploit continuous variables emerge,  it is   natural to investigate quantum entanglement in these systems\cite{content1,content2}.

    We have obtained the exact solution of the \emph{Heisenberg} equations of motion for the \emph{operators} $q_{\pm}$ of the quantum oscillators.

    In the Heisenberg picture the states do not evolve in time and correlations are obtained from expectation values of the Heisenberg operators in the \emph{initial state} as discussed in the previous sections. We have considered a factorized initial state of the system corresponding to the vacuum for each normal mode of the oscillators, independently, namely an uncorrelated initial state.  However, we find that coherence, namely an off-diagonal correlation   mediated by the coupling to the bath emerges and remains in the asymptotic state. This suggests that the independent normal modes become entangled during the time evolution as a consequence of their couplings to the bath.   In the Heisenberg picture where the operators evolve in time but the states do not, the issue of entanglement is not explicit. This section is devoted to understanding the time evolution of \emph{the initial state} and the emergence of entanglement between the oscillators mediated by their couplings to the baths. In particular we   relate the coherence exhibited by the off-diagonal correlation functions $\langle q_+(t) q_-(t')\rangle $ in the Heisenberg picture to the quantum entanglement  between the normal modes that emerges in the \emph{time evolution of the initial state}.

    Although we defer a full study of the time evolution of the full density matrix with thermal baths including a corresponding derivation of the master equation to a forthcoming article\cite{next}, here we consider the simpler case of the two baths at zero temperature.

    Rather than using the Heisenberg picture wherein the full time dependence is in the operators, and the quantum states do not evolve in time, we find that the connection between the coherence and quantum entanglement is best exhibited in the interaction picture. Hence we write eqn. (\ref{renHS}) in terms of
    \be H_0 = H_S + H_B \,,\label{h0} \ee where $H_S$ is diagonal in the renormalized normal mode basis and the interaction Hamiltonian is $H_{SB}$ plus the contribution of the  counterterm matrix, but this latter term does not lead to entanglement. In order to understand the emergence of entanglement it is sufficient to consider the interaction term $H_{SB}$,
    \be H_{SB}(t) = -(q_+(t) b_+(t)+ q_-(t)b_-(t))\,, \label{Hsbint}\ee where in the normal mode basis and using eqn. (\ref{SBNM}) we have introduced
    \bea b_+(t)  & = & \cos(\psi)B_1[\{Q_1(t)\}] - \sin(\psi) B_2[\{Q_2(t)\}] \label{bplus}\\
    b_-(t)  & = & \sin(\psi)B_1[\{Q_1(t)\}] + \cos(\psi) B_2[\{Q_2(t)\}]\,, \label{bmin} \eea  with $B_1[\{Q_1\}],B_2[\{Q_2\}]$   given by eqn. (\ref{baths}) and $\psi$ are the renormalized angles. The time evolution of the operators $q_\pm(t)$ and $Q_{1,2}(t)$ is that of ``free fields'' with $H_0$ as befits the interaction picture.

    We now consider an initial state which is a tensor product of the vacua for the normal modes and for the bath  degrees of freedom, namely
    \be |\Psi^{(0)}(0) \rangle = |0\rangle_S \otimes |0\rangle_{B}~~;~~ |0\rangle_S=|0\rangle_{+}\otimes |0\rangle_{-}~~;~~ |0\rangle_B=|0\rangle_{B1}\otimes |0\rangle_{B2}\, \label{psi0}\ee where $|0\rangle_{B1},|0\rangle_{B2}$  stand    for the vacuum states for the independent bath oscillators. We obtain the time evolved state up to second order in perturbation theory in the system-bath coupling (in the normal mode basis) to highlight the emergence of entanglement between the normal modes $q_\pm$ as a consequence of their coupling to the baths. We postpone a non-perturbative and non-zero temperature discussion to a forthcoming study.

    Evolving the initial state (\ref{psi0}) with the unitary time evolution operator in the interaction picture we find
    \be |\Psi(t)\rangle = |\Psi^{(0)}(0)\rangle + |\Psi^{(1)}(t)\rangle + |\Psi^{(2)}(t)\rangle +\cdots \label{psioft}\ee where
    \bea |\Psi^{(1)}(t)\rangle & = &  +i\int^t_0 dt_1 \Bigg\{\frac{e^{i\Omega_{R+} t_1}}{\sqrt{2\Omega_{R+}}}\,\Big(|1\rangle_+ \otimes|0\rangle_-\Big)\, b_+(t_1)|0\rangle_B  \nonumber \\  & +&
    \frac{e^{i\Omega_{R-} t_1}}{\sqrt{2\Omega_{R-}}}\,\Big(|0\rangle_+ \otimes |1\rangle_- \Big)\, b_-(t_1)|0\rangle_B \Bigg\}\,.\label{psi1}\eea The state up to first order $(|\Psi^{(0)}(0)\rangle + |\Psi^{(1)}(t)\rangle) $  displays entanglement between   the normal modes of the system and bath degrees of freedom.


    Using the relations (\ref{bplus},\ref{bmin}) we can rewrite this state in a manner that displays explicitly the entanglement between the normal modes and the normal modes and the baths, namely
    \bea |\Psi^{(1)}(t)\rangle & = &  +i\int^t_0 dt_1 \Bigg\{\Bigg[\cos(\psi)\,\frac{e^{i\Omega_{R+} t_1}}{\sqrt{2\Omega_{R+}}}\,\Big(|1\rangle_+ \otimes|0\rangle_-\Big)+ \sin(\psi)\,  \frac{e^{i\Omega_{R-} t_1}}{\sqrt{2\Omega_{R-}}}\,\Big(|0\rangle_+ \otimes |1\rangle_- \Big)\Bigg]\, B_1[\{Q_1(t_1)\}]|0\rangle_B  \nonumber \\  & +&\Bigg[-\sin(\psi)\,\frac{e^{i\Omega_{R+} t_1}}{\sqrt{2\Omega_{R+}}}\,\Big(|1\rangle_+ \otimes|0\rangle_-\Big)+ \cos(\psi)\,  \frac{e^{i\Omega_{R-} t_1}}{\sqrt{2\Omega_{R-}}}\,\Big(|0\rangle_+ \otimes |1\rangle_- \Big)\Bigg]\, B_2[\{Q_2(t_1)\}]|0\rangle_B  \Bigg\}\,.\label{psi1ent}\eea The terms in the brackets multiplying the individual bath operators are entangled states of the two oscillators, and the full state is recognized as entangled states of the oscillators entangled with the baths. This first order state shows the emergence of entanglement between the oscillators mediated by their couplings to the baths explicitly as a function of time.


    The second order contribution   is given by\footnote{In principle the second order contribution should include the counterterm matrix in eqn. (\ref{renHS}). However such contribution will only cancel terms that diverge with the bandwidths but do not lead to an entangled state upon tracing the degrees of freedom of the baths.  }

    \be |\Psi^{(2)}(t)\rangle = +i \int^t_0 dt_2 \Big(q_+(t_2) b_+(t_2)+ q_-(t_2)b_-(t_2)\Big)|\Psi^{(1)}(t_2)\rangle\,. \label{psi2} \ee There are several contributions to this entangled state of bath and mechanical oscillators degrees of freedom; for example, the term $q_-(t_2) b_-(t_2)$, when applied to the first term on the right hand side of (\ref{psi1}), yields the state $|1_+\rangle \otimes |1_-\rangle \otimes b_-(t_2) b_+(t_1) (|0\rangle_{B1}\otimes |0\rangle_{B2})$. This term contributes to the coherence in second order as discussed below.

    The reduced density matrix for the system is obtained as
    \be \rho_S(t) = \mathrm{Tr}_{B} \Big(|\Psi(t)\rangle\langle \Psi(t)|\Big) \label{rhored}\ee where $\mathrm{Tr}_{B}$ is the trace over the degrees of freedom of the baths $Q_1,Q_2$; it is straightforward to show that
    \be \langle \Psi(t)|q_\pm(t) |\Psi(t) \rangle = \mathrm{Tr}_S \Big(\rho_S(t)\, q_\pm(t)   \Big) = 0 \,. \label{onept}\ee

   \noindent  The coherence is obtained as
    \be \langle \Psi(t)|q_+(t) q_-(t)|\Psi(t) \rangle = \mathrm{Tr}_S \Big(\rho_S(t)\, q_+(t) q_-(t)  \Big) \,. \label{coherho}\ee Up to second order the coherence is obtained by expanding the state $|\Psi(t)\rangle$ as in (\ref{psioft}), and  it is straightforward to confirm that the zeroth and first order contributions vanish. Therefore up to  second order we find
    \be \langle \Psi(t)|q_+(t) q_-(t)|\Psi(t) \rangle =  \Big[\langle \Psi^{(0)}(0)|q_+(t) q_-(t)|\Psi^{(2)}(t)\rangle + \langle \Psi^{(2)}(t)|q_+(t) q_-(t)|\Psi^{(0)}(0)\rangle + \langle \Psi^{(1)}(t)|q_+(t) q_-(t)|\Psi^{(1)}(t)\rangle\Big]\,. \label{2ndord} \ee

    The contribution from the last term in (\ref{2ndord}) is obtained readily by noticing that
    \bea q_+(t) q_-(t) \,\big(|1\rangle_+ \otimes |0\rangle_-\big)   &     \propto  & \big(|0\rangle_+ \otimes |1\rangle_-\big) +\cdots \nonumber \\
    q_+(t) q_-(t)\,\big( |0\rangle_+ \otimes |1\rangle_- \big)  & \propto &   \big(|1\rangle_+ \otimes |0\rangle_-\big) +\cdots \,, \label{qqpsi1}\eea where the dots stand for terms with two excitations. Therefore the overlap in the last term in (\ref{2ndord}) is determined by $_{B}\langle b_+(t_2)b_-(t_1)\rangle_B $.

    In the first two terms the contribution to  the coherence arises from the two-excitation state  $|1\rangle_+ \otimes |1\rangle_-$ (in obvious notation) in $|\Psi^{(2)}(t)\rangle$, because $q_+(t) q_-(t)|0\rangle_+\otimes |0\rangle_- \propto |1\rangle_+ \otimes |1\rangle_- $. This term in $|\Psi^{(2)}(t)\rangle$ arises from $q_-(t_2) b_-(t_2)$ applied to the first term and
    $q_+(t_2) b_+(t_2)$ applied to the second term
    in (\ref{psi1}). Therefore the first and second overlaps in (\ref{2ndord}) are also determined by $_{B}\langle b_+(t_2)b_-(t_1)\rangle_B $. The second order contribution to the coherence is   determined by
    \be _{B}\langle 0| b_+(t_1)b_-(t_2)|0\rangle_B   = \sin(\psi)\cos(\psi)\frac{1}{\pi}\int_{-\infty}^\infty\big[\sigma_1(\omega')-\sigma_2(\omega') \big]\,\Theta(-\omega')\,e^{i\omega'(t_1-t_2)}\,d\omega'\label{bathcorre2nd}\ee (integrated over $t_1,t_2$ with various exponentials of the normal mode frequencies) where   the interaction picture expansion (\ref{zeroQ1},\ref{zeroQ2}) and the spectral representations (\ref{spec1},\ref{spec2}) have been used. The terms in the bracket  in (\ref{bathcorre2nd}) yield  the noise correlation functions (\ref{correnoise}) in the zero temperature limit, with $n_i(\omega') \rightarrow \Theta(-\omega')$. Furthermore, it is clear that  the coherence vanishes   when the spectral densities of both baths are the same, in agreement with the results obtained in the previous sections when the baths feature the same temperature, and spectral densities as in the weak coupling case.

     A direct calculation of the integrals in the Ohmic case yields the result
    \be \langle \Psi(t)|q_+(t) q_-(t)|\Psi(t) \rangle \propto (\gamma_1-\gamma_2)\,t^2 \label{timev}\ee which quantifies the early time evolution of entanglement and coherence.

  A corollary of this discussion is that the emergence of coherence is a consequence of the entanglement between the system and bath degrees of freedom, which upon tracing over the bath induces entanglement between the two normal modes leading to the non-vanishing off-diagonal correlation function of $q_+$ and $q_-$, that is, the coherence. This perturbative argument while revealing the emergence of entanglement and the phenomenon of coherence mediated by the coupling to the baths cannot capture the full non-equilibrium dynamics. However, is broadly in agreement with the observation in ref.\cite{buttner} in that the coupling to the bath(s) mediates entanglement between the independent normal modes of the mechanical oscillators. A non-perturbative assessment of the asymptotic dynamics of entanglement requires the  quantum master equation; such study will be the focus of a forthcoming study\cite{next}.

    \section{Discussion}\label{subsec:discussion}
    The are several aspects of the results obtained above that merit further discussion.

    \begin{itemize}
    \item{We have focused on the one   and two point correlations functions only simply because the theory is Gaussian, and, therefore higher order correlation functions can be obtained from Wick's theorem. With vanishing initial conditions for the one point functions, correlation functions of odd number of oscillator coordinates will vanish.  However, whereas the two point correlation functions are a sum of the contribution from the initial conditions and bath correlations (see discussion following eqn. (\ref{split})), higher order correlation functions will feature factorized products with mixed contributions from initial conditions and noise correlators.  In particular for the case of vanishing detuning when both oscillators couple only to one bath, the correlation functions will retain memory of the initial conditions as a consequence of the undamped mode. }

    \item{In the case of a strictly Ohmic baths the coincidence limit of the two point correlation function of the canonical momenta exhibits a logarithmic dependence on the bandwidths of the baths, namely,  $\langle \langle p^2_\pm (t) \rangle \rangle \propto \ln[\Lambda_j] $ which as shown above, is the same as in the zero temperature limit. This dependence survives in the asymptotic long time regime when the system becomes stationary and is a consequence of the fact that probing correlations on short time scales implies probing high frequency components of the spectral density of the baths.  Although we focused on the strictly Ohmic case because it allows a complete analytic study, the short time divergences are of a more general nature, in particular we expect that in a super-Ohmic case the short time singularities will yield stronger dependence on the bandwidths of the baths. Furthermore, correlation functions of higher time derivatives of the system coordinates will also feature stronger dependence on the bandwidths, suggesting a breakdown of universality in the coincidence limit of these correlation functions. The  physical reason behind these ``divergences'' is clear: Probing correlation functions at nearby time intervals probes the high frequency components and higher moments of the spectral densities of the baths which are very sensitive to the cutoff functions. The analysis above suggests that the near coincidence limit of the correlation functions in the stationary regime must be defined by separations in time  much larger than the inverse cutoff frequency of the baths but much smaller than the inverse (renormalized) frequencies of the mechanical oscillators and relaxation times, in other words, $\tau \gg 1/\Lambda_j$ but $\tau \ll \Omega_\pm, \gamma_j$. }

     \item{The perturbative analysis of section (\ref{sec:entanglement}) establishes a correspondence between the off-diagonal coherence $\langle \langle q_+(t) q_-(t) \rangle \rangle$ and entanglement between the normal modes of the mechanical oscillators mediated by their couplings to the baths. While a non-perturbative correspondence with finite temperature baths will be discussed elsewhere, our conclusion is broadly consistent with that of ref.\cite{buttner}. However, our analysis and results differ from those in this reference in various important aspects.  Whereas ref.\cite{buttner} finds that initial two-mode entanglement does not survive in the stationary state when coupling to two baths, we find that when the normal modes of the mechanical oscillators are initially uncorrelated, their couplings to the baths induces   coherence that not only \emph{survives} in the stationary regime but also in the high temperature (classical) regime \emph{if} the temperatures of the baths   are \emph{different}.  Although we have established the survival of  the coherence at high temperature, and   established a correspondence between coherence and entanglement at zero temperature in perturbation theory, it is physically reasonable to conclude that  if the normal modes are initially uncorrelated, entanglement between the normal modes of the mechanical oscillators mediated by their couplings to the baths would be the only explanation for a non-vanishing coherence in the stationary regime. This suggests, more generally,  that entanglement will also survive both in the stationary and the high temperature regimes under the same circumstances.  Our results for the case of both mechanical oscillators coupling to a single bath, also studied in \cite{keeling}, are broadly consistent with the results of this reference in that for vanishing detuning there is an undamped mode. However, we do not find any singularity in the correlation functions as the detuning vanishes, and we find asymptotically at long times a finite, constant contribution to the coherence   from initial conditions that reflects these conditions. For small detuning we find   that the undamped mode obtains a very small damping (relaxation) rate, and at asymptotically long times the contribution from the initial conditions vanishes. This is expected:  For any non-vanishing damping rate for the \emph{two} modes, however small,   memory of the initial conditions will vanish at asymptotically long time. It is only in this sense that there is a discontinuity in the case of vanishing detuning, a natural consequence of the non-commutativity of the long time and vanishing detuning limits, but otherwise we do not find any singularity  in the correlation functions.  }

    \end{itemize}

\section{Conclusions and further questions:}
In this article we have studied the non-equilibrium dynamics of two coupled mechanical oscillators with general couplings to two uncorrelated baths with different spectral densities and temperatures. Our study is motivated by recent advances in cavity electrodynamics and optomechanics as continuous-variables platforms for quantum computing and information, and the experimental possibility to engineer the environmental degrees of freedom  to control decoherence and dissipation in these systems. We obtained the general solution of the Heisenberg-Langevin equations as an initial value problem to understand the time evolution of correlation functions towards an asymptotic stationary state. The normal modes of the  mechanical oscillators ``mix'' through their coupling to the different baths, and this phenomenon leads to the emergence of coherence (off diagonal correlation functions)   when the normal modes of the mechanical oscillators are uncorrelated in the initial state. The mixing of the mechanical degrees of freedom mediated by the baths introduces novel renormalization aspects which are discussed in detail. The case of Ohmic baths with different spectral densities and temperatures provides an arena for an exact analytic treatment, and we focus on two relevant limits: a strong coupling regime that includes the case of vanishing detuning of the normal modes of the mechanical oscillators, and a weak coupling regime characterized by weak damping and dissipation. In both cases we find that a \emph{non-equilibrium stationary state} emerges asymptotically at long times for $T_1\neq T_2$. We obtain the complete expression for the unequal time correlation functions both for coordinates (normal modes) and their canonical momenta in the strong and weak coupling regimes.  We find that the coherence (off-diagonal correlation functions) \emph{survives} in the non-equilibrium case for the  high temperature  (classical) limit of both baths if the baths feature different spectral densities and/or temperatures.


  The main physical explanation behind the survival of coherence, an unexpected and perhaps counterintuitive result,  is   the fact that the off-diagonal correlations in the normal mode basis are related to the \emph{difference} of the correlation functions of bath variables. As a result, the coherence survives in the high temperature    limit if  the baths feature different   temperatures but is strongly suppressed if the baths feature the \emph{same} temperature, vanishing exactly if the baths feature \emph{both} the same spectral density \emph{and} temperatures. This important aspect is also confirmed by the perturbative study of the emergence of normal mode entanglement.

  In the high temperature but equilibrium case, $T_1 = T_2 \equiv T$, with $T$ much larger than the normal mode frequencies, the asymptotic long time averages of normal mode  coordinates obey classical equipartition, and the coherence is   suppressed both in the strong and weak coupling cases.     Therefore, the survival of coherence in the high temperature regime is \emph{inherently} a consequence of the non-equilibrium nature of the asymptotic stationary state for $T_1\neq T_2$.

In the case of weak couplings,  implying that the relaxation rates are much smaller than the typical frequencies of the normal modes, the contribution to the coherence from initial conditions leads to interference beats that may be observable in an intermediate time regime.

 We discussed the dependence of the correlation functions on the bandwidths of the baths arguing that for the low energy (frequency) effective description to be insensitive to the high frequency details of the baths,  the coincidence limit of the correlation functions in the stationary state must be interpreted carefully. The particular case of vanishing detuning between the normal modes of the mechanical oscillators along with having both oscillators   coupled to only one bath is  peculiar in the sense that the degeneracy leads to an undamped mode, and the asymptotic long time limit of the correlation functions retain information on the initial conditions. This \emph{may} signal the unsuitability  of  a Markovian approximation to the dynamics within the framework of the quantum master equation for the reduced density matrix.

We provide a perturbative analysis that establishes a correspondence between the off-diagonal coherence and entanglement between the normal modes of the mechanical oscillators \emph{mediated} by their couplings to the different baths. This perturbative analysis, carried out in the limits of zero temperature of the baths, only captures the early time dynamics of entanglement and cannot address the fully non-perturbative nature of the asymptotic stationary state. Nevertheless,  it allows us to argue that if the normal modes of the mechanical oscillators are uncorrelated in the initial state, the emergent  coherence is a result of the entanglement between the normal modes \emph{mediated by their couplings to the baths}. This analysis leads us to   \emph{conjecture}   that the survival of the coherence in the high temperature (classical) limit found above  entails the concomitant survival of entanglement between these degrees of freedom mediated by the baths in the that limit and is \emph{inherently} a consequence of the non-equilibrium asymptotic state.

This conclusion suggests that designing an experimental setup with mechanical oscillators coupled to two baths at different temperatures may provide a platform to maintain coherence and entanglement in the high temperature limit.

The result of this analysis suggests further questions. The exact solution reveals several subtle aspects   such as  the non-equilibrium nature of the asymptotic stationary state in the case of different bath temperatures and/or couplings and the persistence of memory of the initial conditions in the case of vanishing detuning with both oscillators coupled to only one bath. These subtle aspects will require a careful derivation of a quantum master equation, that would correctly account for the ``mixing'' between the normal modes and the off-diagonal energy shifts. The current study provides the groundwork for a consistent derivation of such quantum master equation.   Further investigation  of these aspects will be reported elsewhere.

\acknowledgements The authors thank X.-L. Wu for fruitful and illuminating discussions. D. B. gratefully   acknowledges support from NSF through grant PHY-1506912. D.J. gratefully acknowledges the continued support of the Dietrich School of Arts and Sciences of the University of Pittsburgh.

\appendix
\section{Functions $F_j(\tau)$:} \label{app:F}
 In the strictly Ohmic limit with $\mathrm{Im}\chi_j(\omega) = \gamma_j \, \omega$,  the   integrals in eqn. (\ref{Fdefs}) can be carried out   by using the identities
 \be n(\omega) = -\frac{1}{2}+\frac{1}{2}\,\coth[\frac{\omega}{2T}]~~;~~\coth[z/2] = \frac{2}{z} + \sum_{l=1}^\infty \frac{4z}{(2\pi l)^2 +z^2} \label{coth}\ee featuring simple poles in the upper and lower half complex $\omega$ plane. We consider $\tau >0$ selecting the poles in the upper half plane. We find
 \be F_j(\tau) = F^{(a)}_j(\tau)+ F^{(b)}_j(\tau)\label{f1f2}\ee with
 \be F^{(a)}_j(\tau)  = \frac{1}{2W_j}~\frac{e^{-\gamma_j\tau/2}}{2} \Bigg[e^{iW_j\tau}\,\coth \Big[\frac{W_j+ i\frac{\gamma_j}{2}}{2T_j}\Big]+e^{-iW_j\tau}\coth\Big[\frac{W_j- i\frac{\gamma_j}{2}}{2T_j}\Big]-i\sin[W_j \tau]  \Bigg]  \,, \label{F1js}\ee and
 \be F^{(b)}_j(\tau)= -\frac{\gamma_j}{4\pi^3\,T^2_j}~\sum_{l=1}^{\infty} \frac{e^{-2\pi l T_j\tau}}{l^3 \Bigg[ \Big(1+\big(\frac{W}{2\pi l T_j}\big)^2 \Big)^2 - \Big( \frac{\gamma_j}{2\pi l T_j}\Big)^2 \Bigg]}\,.  \label{F2js}\ee

 \section{Functions $J^{j}_{\alpha\beta}(\tau)$}\label{app:J}
The functions $J^{j}_{\alpha\beta}(\tau), \alpha,\beta = \pm$ in eqns. (\ref{qppWC}-\ref{qpmWC}) are given by

     \bea J^{(j)}_{\pm \pm}(\tau)  & = &  \int^{\infty}_{-\infty} \frac{d\omega}{\pi}\,  \frac{\omega\,e^{i\omega \tau }\,\gamma_j n_j(\omega)}{\Big[\big(\omega-i \frac{\Gamma_\pm}{2}\big)^2 -\Omega^2_\pm \Big]\Big[\big(\omega+i \frac{\Gamma_\pm}{2}\big)^2 -\Omega^2_\pm\Big]}\label{Jjpmpm}\\
     J^{(j)}_{+-}(\tau) & = & \int^{\infty}_{-\infty} \frac{d\omega}{\pi}\,  \frac{\omega\,e^{i\omega \tau }\,\gamma_j n_j(\omega)}{\Big[\big(\omega-i \frac{\Gamma_+}{2}\big)^2 -\Omega^2_+ \Big]\Big[\big(\omega+i \frac{\Gamma_-}{2}\big)^2 -\Omega^2_-\Big]} ~~;~~ j=1,2 \,.\label{Jjpm}\eea
     In the limit of infinite bandwidth these integrals are finite and can be carried out by residues  using the relations  (\ref{coth}). For $\tau   > 0$ we find that the   functions $J^{(j)}_{\alpha \beta}, \alpha, \beta = +,-$   above can be written respectively as $J^{(j,a)}_{\alpha \beta} + J^{(j,b)}_{\alpha \beta}$ where to leading order in the ratios $\Gamma_\pm /\Omega_\pm$ we find for $\tau >0$
     \be J^{(j,a)}_{\pm\pm}(\tau) =  \frac{\gamma_j}{2\,\Gamma_\pm\,\Omega_\pm}\,e^{-\Gamma_\pm \tau/2} \, \Bigg[-i  \sin(\Omega_\pm\tau)  + \cos(\Omega_\pm\tau) \coth\Big[\frac{\Omega_\pm}{2T_j} \Big]    \Bigg] \label{Jppma} \ee
 \be J^{(j,b)}_{\pm\pm}(\tau) =  -\frac{\gamma_j}{4\pi^3 T^2_j} \sum_{l=1}^\infty \frac{e^{-2\pi l T_j \tau}}{l^3\,\Big[\big(1-\frac{\Gamma_\pm}{4\pi l T_j} \big)^2+\big(\frac{\Omega_\pm}{2\pi l T_j} \big)^2 \Big]\Big[\big(1+\frac{\Gamma_\pm}{4\pi l T_j} \big)^2+\big(\frac{\Omega_\pm}{2\pi l T_j} \big)^2 \Big]} \label{Jppmb} \ee

  \bea J^{(j,a)}_{+-}(\tau) & = &  \frac{\gamma_j}{2\,(W\Delta)^2}~ e^{-\Gamma_+ \tau/2} \, \Bigg[ i (W\Delta) \cos(\Omega_+\tau) -i \frac{\gamma_1+\gamma_2}{2} \Omega_+  \sin(\Omega_-\tau)   \nonumber \\ & + &  \coth\Big[\frac{\Omega_+}{2T_j}\Big]\Big( \frac{\gamma_1+\gamma_2}{2}\,\Omega_+\,\cos(\Omega_+\tau)\, +W\Delta \, \sin(\Omega_+\tau)\Big)\Bigg] \label{Jpma} \eea
 \be  J^{(j,b)}_{+-}(\tau)   =    -\frac{\gamma_j}{4\pi^3 T^2_j} \sum_{l=1}^\infty \frac{e^{-2\pi l T_j \tau}}{l^3\,\Big[\big(1-\frac{\Gamma_+}{4\pi l T_j} \big)^2+\big(\frac{\Omega_+}{2\pi l T_j} \big)^2 \Big]\Big[\big(1+\frac{\Gamma_-}{4\pi l T_j} \big)^2+\big(\frac{\Omega_-}{2\pi l T_j} \big)^2 \Big]}\,. \label{Jpmb} \ee

\end{document}